\documentclass[pre,aps]{revtex4-2}

\usepackage{amsmath, amssymb, amsfonts, mathrsfs}
\usepackage{graphicx, color, xcolor}
\usepackage{bbm, dsfont}
\usepackage{enumitem}
\usepackage{comment}

\usepackage{amsthm}
\newtheorem{theorem}{Theorem}[section]
\newtheorem{corollary}[theorem]{Corollary}
\newtheorem{lemma}[theorem]{Lemma}
\newtheorem{proposition}[theorem]{Proposition}

\newtheorem{definition}[theorem]{Definition}
\newtheorem{example}[theorem]{Example}

\newtheorem{claim}[theorem]{Claim}

\numberwithin{equation}{section}

\newcommand{\f}[2]{\frac{#1}{#2}}
\newcommand{\p}{\partial}

\renewcommand{\Re}{\mathrm{Re}}

\newcommand{\vertiii}[1]{{\left\vert\kern-0.25ex\left\vert\kern-0.25ex\left\vert #1
    \right\vert\kern-0.25ex\right\vert\kern-0.25ex\right\vert}}

\newcommand{\dpr}[2]{\langle #1, #2 \rangle}

\newcommand{\al}{\alpha}

\newcommand{\eps}{\epsilon}

\newcommand{\la}{\lambda}

\newcommand{\si}{\sigma}

\newcommand{\om}{\omega}

\newcommand{\rone}{\mathbb{R}}
\newcommand{\rtwo}{\mathbb{R}^2}

\newcommand{\cd}{\mathcal{D}}

\newcommand{\ch}{\mathcal{H}}
\newcommand{\cj}{\mathcal{J}}

\newcommand{\cl}{\mathcal{L}}
\newcommand{\cm}{\mathcal{M}}

\newcommand{\beq}{\begin{equation}}
\newcommand{\eeq}{\end{equation}}
\newcommand{\beqna}{\begin{eqnarray*}}
\newcommand{\eeqna}{\end{eqnarray*}}
\newcommand{\beqn}{\begin{equation*}}
\newcommand{\eeqn}{\end{equation*}}
\newcommand{\bp}{\begin{proof}}
\newcommand{\ep}{\end{proof}}
\newcommand{\bprop}{\begin{proposition}}
\newcommand{\eprop}{\end{proposition}}
\newcommand{\bt}{\begin{theorem}}
\newcommand{\et}{\end{theorem}}
\newcommand{\bex}{\begin{example}}
\newcommand{\eex}{\end{example}}
\newcommand{\bc}{\begin{corollary}}
\newcommand{\ec}{\end{corollary}}
\newcommand{\bcl}{\begin{claim}}
\newcommand{\ecl}{\end{claim}}
\newcommand{\bl}{\begin{lemma}}
\newcommand{\el}{\end{lemma}}

\begin{document}

\title{On a Klein-Gordon Reduction for Oscillons}

\author{Atanas G. Stefanov}
\affiliation{Department of Mathematics, University of Alabama - Birmingham,
1402 10th Avenue South, Birmingham AL 35294, USA}

\author{Milena Stanislavova}
\affiliation{Department of Mathematics, University of Alabama - Birmingham,
1402 10th Avenue South, Birmingham AL 35294, USA}

\author{Jes\'us Cuevas-Maraver}
\affiliation{Grupo de F\'{\i}sica No Lineal, Departamento de F\'{\i}sica Aplicada I,
Universidad de Sevilla, Escuela Polit\'{e}cnica Superior, C/ Virgen de \'{A}frica, 7, 41011 Sevilla, Spain}
\affiliation{Instituto de Matem\'{a}ticas de la Universidad de Sevilla (IMUS), Edificio
Celestino Mutis, Avda. Reina Mercedes s/n, 41012 Sevilla, Spain}

\author{Panayotis G. Kevrekidis}
\affiliation{Department of Mathematics and Statistics, University
of Massachusetts, Amherst, Massachusetts 01003-4515, USA}
\affiliation{Department of Physics, University
of Massachusetts, Amherst, Massachusetts 01003, USA}

\begin{abstract}
In the present work we examine the dynamics of a model
for oscillons in 1-dimensional space-time field theories
with a cubic nonlinearity. We utilize a reduction of
the model to first and third harmonics, which leads
to a reduced partial differential equation (PDE) system whose
steady states are candidates for the original PDE oscillons.
We analyze the steady states of this model and their 
stability, including via tools such as index theory.
We develop suitable functionals needed for the study of
such stationary states, as well as an analogue of the
famous Vakhitov-Kolokolov criterion for a quantity whose
change of monotonicity reflects a change of stability.
Then, we test the relevant predictions,
{ over the full range
of oscillon frequencies,} through systematic
numerical computations of both the reduced model, its steady
states and stability, and also of the original PDE model, identifying 
its time-periodic oscillon solution. Our results yield some significant
connections with previous studies, but also some fundamental
new insights both on the reduced system and  the
dynamics of the original system.
\end{abstract}

\maketitle

  \section{Introduction}

The study of oscillons is a topic of wide interest, notably
in the context of cosmological models and their
inflationary dynamics~\cite{doi:10.1142/S0218271807009954}.
Such solutions have appeared in the context of 
Einstein-Klein-Gordon equations~\cite{Kou_2021}, in the
study of the dynamics of strings~\cite{Antusch2018},
as well as in the analysis and computation of electroweak
interactions~\cite{Graham1,Graham2}. The relevant states
have also been shown to be of interest in the context
of two-dimensional models of the Abelian-Higgs type~\cite{prdvassos}. In such settings, they have
also been spontaneously detected, e.g., as remnants
of vortex-antivortex annihilations~\cite{PhysRevD.76.041701}.
Finally, they have been argued to be relevant to three-dimensional
variants of the famous $\phi^4$-model~\cite{PhysRevD.76.041701}.
Indeed, they generally constitute a pillar of solitonic 
features of relevance to a wide range of models
in areas of field theory and cosmology~\cite{Rajaraman1982,Manton2004,Weinberg2012,Vilenkin1994}. It is worth noting here also that in other settings 
including dissipative systems and most notably
vibrating granular beds~\cite{PhysRevE.53.2972}, oscillon
waveforms have been observed experimentally~\cite{Umbanhowar1996}.

A significant number of studies has focused on the
classic $\phi^4$ oscillon~\cite{copeland}, which lives on top
of a unit background. Yet, a variant of the problem that 
possesses interesting dynamical features is that with the
``flipped sign'' of the nonlinearity, as considered, 
e.g., early on by Kosevich and Kovalev in~\cite{KosevichKovalev1975}. The latter model
has a vanishing field as the stable homogeneous equilibrium
on top of which the oscillon lives. The model is also
rather nontrivial to analytically study, among other reasons,
due to its potential blow up singularity features~\cite{ACHILLEOS20131}. In a recent
development, the work of~\cite{baras_oscillon} provided
a systematic approach towards the needed multiple scale
nature of a variational formalism, so as to capture
the proper collective coordinate dynamical description
of the oscillon and its potential regime of dynamical
stability. Another key recent study identified
the diagram of the energy vs. frequency of the 
oscillon structures, albeit in the regular $\phi^4$ model
in the work of~\cite{Alexeeva2024}.

In the present work, we take a complementary approach
to the ones utilized so far, as a method that can be
applied to this problem, but also could be of interest
to other settings with spatially localized, temporally
periodic solutions. In particular, in the spirit of
a few mode Galerkin approach, we decompose the solution into
the fundamental mode and its third harmonic (due to 
the nature of our nonlinearity) and truncate higher
order modes. In this way, and ascribing spatio-temporal
dependence, aiming to capture the spatial dependence
and the slow temporal dynamics, we provide a methodology
that can be used to obtain meaningful trial solutions
towards the full problem. I.e., while the full problem
may have ``breather-type'' solutions~\cite{Flach2008,Aubry2006},
we aim to approximate these as steady state (localized) 
spatial
profiles multiplying the first and third harmonics, while
the implicit assumption is that higher order terms may be
small for this approximation to be meaningful.

In what follows, we will first analyze the resulting
partial differential equation (PDE) models for the
prefactor terms of the first and third harmonic. 
We will bring to bear tools from PDEs and dynamical
systems to provide information for the steady states
of the resulting model and its stability. In the process,
we will devise suitable Lyapunov-like functionals that
will provide stability information regarding their 
extremizing states, and we will obtain a 
Vakhitov-Kolokolov~\cite{VakhitovKolokolov1973}-type 
criterion for this system, involving a suitable 
frequency-weighted variant of the mass of the two
modes, whose change of monotonicity will reflect a change
in stability. The use of index theory will provide
a systematic tool for the identification of the
spectral stability of the different states that
will be uncovered. 
Subsequently, this detailed information will be
tested in numerical computations. The different states
will be numerically identified and continued as a function
of the mode frequency. We will see that the few-mode reduction
has some
remarkable successes and also some notable shortcomings
in capturing the full time-periodic oscillon states.
We believe that this analysis offers a valuable PDE
tool in the arsenal of the applied scientist towards obtaining 
and testing information, not only for oscillon but also for
more general such time-periodic, spatially localized states.

Our presentation will be structured as follows.
In section II we will present the model setup.
In section III we provide some basic mathematical 
background results, while in section IV we construct
the waves. Their stability is examined in section V. 
Section VI contains our numerical results, while section VII
summarizes our findings and provides some directions for
future study.

  \section{Model Setup}
We would like to study the following Klein-Gordon system, \cite{baras_oscillon}
\begin{equation}
\label{10} 
 z_{tt}-z_{xx}+4z - 2z^3=0, (t,x)\in \rone_+\times \rone.
\end{equation}
{As is well-known, such a system is Hamiltonian, with 
\begin{equation}
    \label{bar:10} 
    H[z]=\f{1}{2} \int z_x^2+z_t^2 +4 z^2-\f{1}{2} z^4 dx. 
\end{equation}
}
  Specifically, we are interested in the oscillons, which are real-valued, 
  time-periodic {\it approximate solutions} of the form  
  \begin{equation}
  	\label{20} 
  z=u(t,x) e^{i \om t} + \bar{u}(t,x) e^{-i \om t} + v(t,x) e^{3 i \om t} + \bar{v}(t,x) e^{-3 i \om t}
\end{equation}
    In particular, we plug the ansatz \eqref{20} into the KG equation \eqref{10}, and we equate the terms containing $e^{i \om t}, e^{3 i \om t}$ (i.e. we ignore higher order time harmonics $e^{\pm 5 i \om t}, e^{\pm 7 i \om t}, e^{\pm 9 i \om t})$). The resulting system is the following approximate, coupled Klein-Gordon system, in the form 
    \begin{equation}
    	\label{15} 
    	\left\{
    	\begin{matrix}
	u_{tt}+2 i \om u_t - u_{xx}+(4-\om^2) u -( 6 u^2 \bar{u}+6 \bar{u}^2 v +12 u |v|^2)=0 \\
v_{tt}+6 i \om v_t-v_{xx}+(4-9 \om^2) v -(2 u^3+12 |u|^2 v+6 |v|^2 v)=0.
    	\end{matrix}
    	\right.
    \end{equation}
   Now, we are interested in time independent~\footnote{but not necessarily real-valued, as this property was enforced with the ansatz \eqref{20}} solutions of this system $(u,v)=(\phi, \psi)$, which satisfy 
     the elliptic system 
      \begin{equation}
    	\label{30} 
    \left\{ 	\begin{matrix}
    		-\phi''+(4-\om^2) \phi - 
    		 (6\phi^2 \bar{\phi}+6 \bar{\phi}^2 \psi+12 \phi |\psi|^2)=0,  \\
    		-\psi''+(4-9 \om^2) \psi -(2 \phi^3+12 |\phi|^2\psi +6 |\psi|^2\psi) = 0.
    	\end{matrix}\right. 
    \end{equation}
We are looking for real-valued solutions $\phi, \psi$ of \eqref{30}. In such a case,  the approximate solution \eqref{20} is in the form 
\begin{equation}
	\label{500} 
	2\left[ \phi\cos(\om t)+ \psi \cos(3\om t)\right].
\end{equation}
From this point on, our object of interest is the dynamics of the  system \eqref{15}, specifically the stability of its 
stationary solutions $(\phi, \psi)$ of \eqref{30}.

\begin{theorem}[Existence of bell-shaped  oscillons]
	\label{theo:10} 
	Let $|\om|<\f{2}{3}$. Then, the system \eqref{30} has a solution $(\phi, \psi)$, where both $\phi, \psi$ are real-valued and bell-shaped functions\footnote{positive, even and decreasing on $(0, +\infty)$}. In addition, $\phi, \psi$ are smooth and exponentially localized. More specifically, 
	$$
	 \phi(x) \sim  e^{-\sqrt{4-\om^2}|x|}, \ \ \psi(x)\sim e^{-\sqrt{4-9 \om^2}|x|}, |x|>>1.
	$$
\end{theorem}


\subsection{Linearization around the steady states}
Let us now consider the linearized problem associated with  the system \eqref{15}. To this end, we adopt the variables 
$$
u=\phi+p+i q, v=\psi+r+i s
$$
Plugging this ansatz in \eqref{15}, taking real and imaginary parts separate and ignoring all quadratic and higher order terms leads to 
 \begin{equation}
	\label{200} 
	\left\{
	\begin{matrix}
		p_{tt}-2\om q_t-p_{xx}+(4-\om^2) p -[(18 \phi^2+12 \psi^2+12\phi\psi)p+(6 \phi^2+24 \phi\psi)r]=0 \\
		q_{tt}+ 2\om p_t-q_{xx}+(4-\om^2) q-[(6\phi^2+12\psi^2-12\phi\psi)q+6 \phi^2 s]=0 \\ 
		r_{tt}-6 \om s_t -r_{xx}+(4-9 \om^2) r -[(12\phi^2+18 \psi^2)r+(6\phi^2+24 \psi^2) p]=0 \\
		s_{tt}+6 \om r_t-s_{xx} +(4-9\om^2) s - [6 \phi^2 q +(12\phi^2+6 \psi^2)s]=0.
	\end{matrix}
	\right.
\end{equation}
This is the form of the linearized problem. We now need to convert this to a more standard, first order in time, 
eigenvalue problem. This is done in a number of steps. 
\subsubsection{The eigenvalue problem}
In matrix form, 
\begin{equation}
	\label{218} 
	\left(\begin{matrix}
		p \\q \\ r\\s
	\end{matrix}\right)_{tt}+ \left(\begin{matrix}
	 0 & -2\om & 0 & 0 \\
	 2\om &0 & 0 & 0 \\
	 0&0& 0& -6\om \\
	 0 & 0 & 6\om & 0
\end{matrix}\right) 	\left(\begin{matrix}
p \\q \\ r\\s
\end{matrix}\right)_{t} + \ch 	\left(\begin{matrix}
p \\q \\ r\\s
\end{matrix}\right)=0,
\end{equation}
where 
 \begin{equation}
 	\label{212} 
 	\ch:=\left(\begin{matrix}
 		\ch_1& 0 &   - 24 \phi\psi  -6\phi^2  & 0 \\
 		0 & 	\ch_2  &0  & -6 \phi^2\\
 		- 24 \phi\psi  -6\phi^2  & 0 &\ch_3 & 0\\
 		0 &  -6 \phi^2 & 0 & \ch_4
 	\end{matrix}\right)
 \end{equation}
and 
\begin{eqnarray*}
	\ch_1 &=& -\p_{xx}+(4-\om^2) - 18 \phi^2-12\psi^2- 12 \phi\psi; \ \ 
	\ch_2  = -\p_{xx}+(4-\om^2) - 6 \phi^2 - 12 \psi^2 +12 \phi\psi  \\ 
	\ch_3 &=& -\p_{xx}+(4-9 \om^2) -12 \phi^2  - 18 \psi^2; \ \ 
	\ch_4  =  -\p_{xx}+(4-9 \om^2) -12 \phi^2  - 6 \psi^2
\end{eqnarray*}
Note that the operator $\ch$ is self-adjoint, for nice enough $\phi, \psi$, when taken  with the  domain $H^2(\rone)\times H^2(\rone)\times H^2(\rone)\times H^2(\rone)$, 
while 
$$
\cj:=\left(\begin{matrix}
	0 & -2\om & 0 & 0 \\
	2\om &0 & 0 & 0 \\
	0&0& 0& -6\om \\
	0 & 0 & 6\om & 0
\end{matrix}\right) , 
$$
is skew-symmetric, i.e. $\cj^*=-\cj$. The corresponding eigenvalue problem, which can be obtained by the assignment 
$$
	\left(\begin{matrix}
	p \\q \\ r\\s
\end{matrix}\right)\to e^{\la t} 	\left(\begin{matrix}
p \\q \\ r\\s
\end{matrix}\right)=:e^{\la t}\vec{z}
$$
can be written in the form 
$$
\la^2 \vec{z} +\la \cj	\vec{z}+ \ch \vec{z}=0.
$$
This eigenvalue problem is a pencil of order two, and these have been an object of intense investigations. A standard equivalence in a more classical Hamiltonian form is as follows\footnote{Note that the identity operators in the following formula act 
	$Id: {\mathbf C}^4\to  {\mathbf C}^4$.} 
\begin{equation}
	\label{210} 
	\left(\begin{matrix}
		0 & -Id \\ Id & -\cj 
	\end{matrix}\right)	\left(\begin{matrix}
	\ch & 0 \\ 0 & I 
\end{matrix}\right)
	\left(\begin{matrix}
\vec{z} \\ \vec{w} 
\end{matrix}\right)= 	\la \left(\begin{matrix}
\vec{z} \\ \vec{w} 
\end{matrix}\right)
\end{equation}
This prompts the following definition. 
\begin{definition}
	\label{defi:10} 
	We say that the steady state solution $(\phi, \psi)$ of \eqref{30} is spectrally stable, if the eigenvalue problem \eqref{210} does not have non-trivial solutions with $\Re\la>0, \vec{z}\neq 0$
\end{definition}
\subsection{Stability of the waves} 
Next, we address the important issue of symmetries and the corresponding eigenvalues at zero for the linearized problem,  that arise. 
\subsubsection{Symmetries of the system}
We can identify at least two symmetries of the system \eqref{30}, namely a translational and 
(phase) modulational invariance. Specifically, for every solution $(\phi, \psi)$ of \eqref{30}  and arbitrary real $y$, then 
$(\phi(\cdot-y), \psi(\cdot-y))$,  $(e^{i y} \phi, e^{3i y} \psi)$ are solutions as well. The usual differentiation in the parameter $y$, and taking into account the setup of $\ch$,  provides two elements in $Ker[\ch]$. More concretely, 
$$
\left(\begin{matrix} 
	\phi' \\  0 \\ \psi'\\  0
	\end{matrix} \right), \ \ \ \left(\begin{matrix} 
	0 \\ \phi \\ 0 \\   3 \psi 
\end{matrix} \right) \in Ker[\ch].
$$
\subsubsection{Non-degeneracy of the wave pair $(\phi, \psi)$}

We now introduce the notion of a non-degeneracy for the wave pair $(\phi, \psi)$. 
\begin{definition}
	\label{defi:20} 
	We say that the solution $(\phi, \psi)$ is non-degenerate, if there are no more elements of $Ker[\ch]$ other than those already found. Specifically, 
	$$
	Ker[\ch] = span \left[\left(\begin{matrix} 
		\phi' \\  0 \\ \psi'\\  0
	\end{matrix} \right), \ \ \ \left(\begin{matrix} 
		0 \\ \phi \\ 0 \\   3 \psi 
	\end{matrix} \right) \right]
	$$
	In other words, non-degeneracy means that the only zeros in the kernel of $\ch$ are those  generated by the symmetries of the system. 
\end{definition}
We are now ready to state the main stability result. 
\begin{theorem}
	\label{theo:20}
	Let $\om: |\om|<\f{2}{3}$. Assume that the steady state solution $(\phi, \psi)$ of \eqref{15}, constructed in Theorem \ref{theo:10}, is a non-degenerate, smooth function of $\om$. That is, the mapping $\om\to (\phi, \psi)$ is Frechet differentiable. 
	
	Then, it 
	 is \underline{spectrally stable if and only if} the following holds true 
	 $$
	\om  \p_\om \int_{\rone} \left(\phi^2+9 \psi^2\right) dx +  \int_{\rone} \left(\phi^2+9 \psi^2\right) dx <0.
	 $$
	 Equivalently, $\om\to \om  \int_{\rone} \left(\phi^2+9 \psi^2\right) dx$ is decreasing. 
\end{theorem}


\section{Preliminaries}
We use standard notations for the $L^p$ spaces, namely with the norm $\|f\|_p:=\left(\int_{\rone} |f(x)|^p dx\right)^{1/p}$, and the Sobolev spaces $H^s(\rone)$.  The Sobolev embedding guarantees that for all $2<p<\infty$, there is a constant $C_p$, so that $\|u\|_p\leq C_p \|u\|_{H^1}$.

\subsection{Decreasing rearrangements and compactness criteria}

One tool that will be useful in the sequel is the notion of the decreasing rearrangement. For a function $f:\rone\to \rone$, there exists an even, positive,  non-increasing on $(0, +\infty)$ function $f^*$, so that the corresponding level functions coincide, i.e., 
$
d_f(\al)=\{x: |f(x)|>\al\}=\{x: f^*(x)>\al\}=d_{f^*}(\al)
$
Alternatively, a constructive definition may be given as follows
$$
f^*(t)=\inf\left\{s: d_f(s)\leq 2|t| \right\}, \ \ t\in \rone
$$
Either way, we conclude that $\|f\|_p=\|f^*\|_p$.  Additionally, there are the standard inequalities 
$$
\int_{\rone} f(x) g(x) dx\leq \int_{\rone} f^*(x) g^*(x) dx, \int_{\rtwo} f(x-y) g(y) h(x)  dx dy \leq  
\int_{\rtwo} f^*(x-y) g^*(y) h^*(x)  dx dy
$$
The Szeg\"o inequality guarantees that $f^*\in H^1$, whenever $f\in H^1(\rone)$ and moreover, 
\begin{equation}
	\label{60} 
	\|\p_x f\|\geq \|\p_x f^*\|
\end{equation}
In case of equality in \eqref{60}, one concludes that the function $f$ coincides with its rearrangement, i.e. $f=f^*$, and therefore it belongs to a class of functions, usually referred to as bell-shaped.

 The Riesz-Relich-Kolmogorov compactness criteria (or rather a consequence thereof), is the following -  a sequence $\{u_n\}_n$ is compact in $L^p(\rone), 1<p<\infty$, if 
\begin{itemize}
	\item $\sup_n \|u_n\|_{H^1(\rone)}<\infty$
	\item For every $\epsilon>0$, there exists $R=R_\epsilon>0$, so that 
	$$
	\sup_n \int_{|x|>R} |u_n(x)|^p dx<\epsilon. 
	$$
\end{itemize}

\subsection{Basics of the instability index theory }
Consider a Hamiltonian eigenvalue problem in the form 
\begin{equation}
	\label{d:10}
	J L u =\la u
\end{equation}
where $J^*=-J$ and $L^*=L$, with appropriate domains for the composition operator $J L$. Assume in addition that both $J, L$ act on spaces of functions, and they do preserve real-valued functions. The stability for \eqref{d:10} is a fundamental problem in the modern dynamics. Specifically, as in Definition \ref{defi:10}, we say that the problem \eqref{d:10} is stable, if there are no non-trivial solutions of \eqref{d:10} with $\Re\la>0$.   
In order to introduce some relevant notions, we assume that $L$ is semi-bounded from below, that is there exists a constant $C$, so that $L\geq C$ in the sense of quadratic forms. In fact, we assume that its negative subspace $X_-$ is finite dimensional and $P_{X_-} L P_{X_-}$ has only point spectrum, each with finite multiplicity. Denote its Morse index by 
$$
n(L)=dim[X_-]=\#\{\si: \si<0: \si \in \si_{p.p.}(L)\}.
$$
Next, let $k_r$  be  the number of positive eigenvalues of \eqref{d:10},  $k_c$ be the number of quadruplets of eigenvalues with non-zero real and imaginary parts, and $k_i^-$, the number of pairs of purely imaginary eigenvalues with negative Krein-signature.  For a simple pair of imaginary eigenvalues $\pm i \mu, \mu\neq 0$, and the corresponding eigenvector
$\vec{h} = \left(\begin{array}{c}
	h_1  \\ h_2
\end{array}\right) $, the Krein signature, either $\pm 1$ is the following quantity
$
sgn(\dpr{ \mathcal{L}  \vec{h}}{ \vec{h}}).
$
Consider  the generalized kernel of $J L$
$$
gKer(J L)=span[(Ker(J L)^l, l=1, 2, \ldots]  .
$$
Assume that $dim(gKer(J L))<\infty$, although this is, strictly speaking, not required in \cite{LZ}. Select a basis in
$$
gKer(J L)\ominus Ker(L)=span[\eta_j, j=1, \ldots, N].
$$
Introduce $\cd\in \cm_{N\times N}$, 
$$
\cd:=\{\cd_{i j}\}_{i,j=1}^N: \cd_{i j}=\dpr{\cl \eta_i}{\eta_j} .
$$
Then, following \cite{LZ}, we have the following formula, relating the number of
``instabilities'' or Hamiltonian  index of the eigenvalue problem \eqref{d:10} and the Morse indices of the operators $L$ and $\cd$, 
\begin{equation}
	\label{d:20}
	k_{Ham}:=k_r+2 k_c+2k_i^-=n(L)-n(\cd).
\end{equation}
Specifically, if $n(L)=1$, it follows from \eqref{d:20} that
$k_c=k_i^-=0$ and
\begin{equation}
	\label{d:30}
	k_r=1-n(\mathcal{D}).
\end{equation}
 Note that in the case $n(L)=1$, instability occurs exactly when $n(\cd)=0$, while stability occurs whenever $n(\cd)=1$. We formulate this in a corollary as follows. 
 \begin{corollary}
 	\label{cor:20} 
 	Assume that in the eigenvalue problem \eqref{d:10}, we have that $n(L)=1$. Then, the  problem \eqref{d:10}  is spectrally stable \underline{if and only if} $n(\cd)=1$. 
 \end{corollary}


\section{Construction of the waves} 
\label{construct}

We can clearly see that the elliptic system \eqref{30} is   of the form 
\begin{equation}
	\label{35} 
	\left\{  \begin{matrix}
		-\phi''+(4-\om^2) \phi - 
		\f{\p F}{\p \bar{\phi}}=0 \\ 
		-\psi''+(4-9 \om^2) \psi -	\f{\p F}{\p \bar{\psi}}=0 
	\end{matrix}\right. 
\end{equation}
where 
\begin{equation}
	\label{40} 
	F(u,v,\bar{u}, \bar{v})= 3 |u|^4  +2 \bar{u}^3 v+ 12 |u|^2 |v|^2 +2 u^3 \bar{v}+3 |v|^4=3(|u|^4+|v|^4+4 |u|^2|v|^2)+4 \Re \bar{u}^3 v.
\end{equation}
{As such, it supports a Hamiltonian structure in the form 
\begin{equation}
    \label{bar:20} 
    H_1[\phi, \psi]=\f{1}{2}\int  |\phi'|^2+|\psi'|^2+(4-\om^2)|\phi|^2+(4-9\om^2)|\psi|^2 
    + 6(|\phi|^4+|\psi|^4+4 |\phi|^2|\psi|^2)+8 \Re [\bar{\phi}^3 \psi] dx
\end{equation}
Clearly, one obtains $H_1[\phi, \psi]=H[z]$, under the transformation \eqref{500}. While here we mention in
passing
the Hamiltonian structure of the steady state
ODEs, naturally, the relevant transformation can
also be used to obtain the Hamiltonian structure
of the PDEs of the system of Eqs.~(\ref{15}).}
\begin{proposition}
	\label{prop:10} 
	Let $|\om|<\f{2}{3}$. Then, the variational problem 
	\begin{equation}
		\label{50} 
		\left\{
		\begin{matrix}
			J[u,v]=\int_{-\infty}^{+\infty}3(|u|^4+|v|^4+4 |u|^2|v|^2)+4 \Re[\bar{u}^3 v]\to \max \\ 
		I[u,v]=	\int_{-\infty}^{+\infty} (u')^2+(v')^2+(4-\om^2)u^2+(4-9\om^2)v^2 dx=1.
		\end{matrix}
		\right.
	\end{equation}
has a solution $(U, V)$, with $U\geq 0, V\geq 0$, and $U,V$ are both even. In addition, $(U, V)$ satisfy the Euler-Lagrange equation, 
 \begin{equation}
	\label{70} 
	\left\{ 	\begin{matrix}
		-U''+(4-\om^2) U- \f{1}{J_0} 
		(6U^3+12 V^2 U + 6 U^2 V)=0,  \\
		-V''+(4-9 \om^2) V  -\f{1}{J_0} (6 V^3+12 U^2 V +2 U^3) = 0.
	\end{matrix}\right. 
\end{equation}
where we denoted  $J_0:=\sup_{I[u,v]=1} J[u,v]=J[U,V]$. 
\end{proposition}

{\bf Remark:} 
\begin{itemize}
\item 
	The solutions $(\phi, \psi)$ of \eqref{35} may be obtained as $\phi=c_0 U, \psi=c_0 V$, where $c_0$ depends only on $J[U, V]$, and so on $\om$ only. 
	\item One concern is that for some values of $\om, 0<\f{2}{3}-|\om|<<1$ (namely the ones for which $4-9\om^2$ is sufficiently small), we may get the ``semi-simple'' solution $U=0, V=C {\rm sech}(\sqrt{4-9 \om^2}x)$ (for suitably chosen $C$, so that $I[0,V]=1$). This would be a minimizer of \eqref{50} anyway, but trivial. We check numerically, see  Fig.~\ref{fig:maximizer} below, that this indeed happens for $0.5316<\om<2/3$.   	
\end{itemize}

\begin{proof}
 The proof is rather straightforward. To this end, note that by Sobolev embedding $\|u\|_4\leq C\|u\|_{H^1}$. Estimating $J[u,v]$ by H\"older's,  we see  that 
 $$
 J[u,v]\leq C (\|u\|^4_4+\|v\|^4_4)\leq C \|u\|^4_{H^1}+\|v\|^4_{H^1})\leq C I^2[u,v].
 $$
 It follows that $J[u,v]$ is bounded from above, when $u,v$ are under the constraint $I[u,v]=1$. Moreover, by elementary properties of the decreasing rearrangements, we have that 
 \begin{eqnarray*}
 	J[u,v] &=& \int_{-\infty}^{+\infty}3(|u|^4+|v|^4+4 |u|^2|v|^2)+4 \Re \bar{u}^3 v\leq \\
 	&\leq &  \int_{-\infty}^{+\infty}3(|u^*|^4+|v^*|^4+4 |u^*|^2|v^*|^2)+4 (u^*)^3 v^* =J[u^*, v^*].
 \end{eqnarray*}
 In terms of the constraints, using the Szeg\"o inequality, we have 
 $$
 I[u, v]\geq I[u^*, v^*].
 $$
 Assuming for a moment that there holds the strict inequality, $1=I[u, v]>I[u^*, v^*]$, say $I[u^*, v^*]=a<1$, we arrive at a contradiction, since then $u^*, v^*$ satisfy 
 $$
 	\left\{
 \begin{matrix}
 	J[u^*,v^*]\geq J[u, v] \\ 
 	I[u^*, v^*]=a<1
 \end{matrix}
 \right.
 $$
 But since the problem is scale invariant, we have that $I[a^{-1/2} u^*, a^{-1/2} v^*]=1$ satisfy the original  constraint, while 
 $$
 	J[a^{-1/2} u^*,a^{-1/2} v^*]=a^{-2} J[u^*, v^*]\geq a^{-2} J[u, v]>J[u,v],
 $$
 in clear contradiction with the setup of the variational problem \eqref{50}. So, $I[u, v]=I[u^*, v^*]$, whence it follows that $u=u^*, v=v^*$, i.e., the variational problem may be taken to maximize $J$ only on the set of  bell-shaped functions. Using the properties of the bell-shaped functions, and the constraint we have that 
 $$
 u^2(x)\leq \f{1}{2x} \int_{-x}^{x} u^2(y) dy\leq \f{1}{2x} \|u\|^2\leq \f{C_\om}{x}. 
 $$
 and similar for $v$. 
 It follows that, for every $R>0$, 
 $$
 \int_{|x|>R} u^4(x) dx\leq C \int_{|x|>R} x^{-2} dx \leq C R^{-1},
  $$
  and similar for $v$. 
  This, together with the constraint $\|u\|_{H^1}+\|v\|_{H^1}\leq C$, implies by the Riesz-Kolmogorov criteria, that any maximizing sequence for \eqref{50} is in fact compact in $L^4(\rone)$. Thus, starting with a maximizing sequence for \eqref{50}, i.e. 
  $$
  I[u_n, v_n]=1, J[u_n, v_n]\to \sup_{I[u,v]=1} J[u,v]
  $$
 and  after taking a strongly convergent in $L^4$, subsequence $\lim_n \|u_n-U\|_4=0, \lim_n \|v_n-V\|_4=0$, whence 
 $$
  \lim_n J[u_n,v_n]=J[U,V].
  $$
   At the same time, by the lower semi-continuity of the $H^1$ norms (with respect to weak convergence, so  certainly with respect to $L^4$ convergence), 
  $$
  I[U, V]\leq \liminf_n I[u_n, v_n]=1.
  $$
  If in fact, we assume  that $I[U,V]<1$, we obtain the same contradiction as above (the value of $\sup_{I[u,v]=1} J[u,v]$ is bigger than it can possibly be), so it follows that $I[U,V]=1$ (and in fact, one may conclude from this that even stronger convergence holds, namely $\lim_n \|u_n-U\|_{H^1}=0, \lim_n \|v_n-V\|_{H^1}=0$).

 We now derive the Euler-Lagrange equations. Taking $U+\epsilon h_1, V+\epsilon h_2$ for some testing functions $h_1, h_2$, we have that the following scalar function achieves a maximum at $\epsilon=0$, 
 $$
 g(\epsilon)=\frac{J[U+\epsilon h_1, V+\epsilon h_2]}{I^2[U+\epsilon h_1, V+\epsilon h_2]}
 $$
 It follows that $g'(0)=0$, while $g''(0)\leq 0$. Note $J_0>0$. We have 
 \begin{eqnarray*}
& &  	J[U+\epsilon h_1, V+\epsilon h_2] = J_0+\\
 	&+& \epsilon(12\dpr{U^3}{h_1}+12\dpr{V^3}{h_2}+24\dpr{V^2 U}{h_1}+24\dpr{U^2V}{h_2}+12\dpr{U^2V}{h_1}+4\dpr{U^3}{h_2})+O(\eps^2)\\
 	& & I^2[U+\epsilon h_1, V+\epsilon h_2] = 1+ 2\epsilon(\dpr{-U''+(4-\om^2)U}{h_1}+\dpr{-V''+(4-\om^2)V}{h_2})+O(\eps^2).
 \end{eqnarray*}
Writing out the first order in $\epsilon$ terms, arising in $g(\epsilon)$ leads to the formulas 
\begin{eqnarray*}
& &  12\dpr{U^3}{h_1}+12\dpr{V^3}{h_2}+24\dpr{V^2 U}{h_1}+24\dpr{U^2V}{h_2}+12\dpr{U^2V}{h_1}+4\dpr{U^3}{h_2}= \\
 &=& 2 J_0(\dpr{-U''+(4-\om^2)U}{h_1}+\dpr{-V''+(4-\om^2)V}{h_2}). 
\end{eqnarray*}
  Since $h_1, h_2$ are independent increments, we have that $U,V$ are weak solutions of the elliptic system \eqref{70}. 
  Standard elliptic estimates for \eqref{70}  imply that such solutions are in fact $H^\infty(\rone)$, so in particular $U,V$ are $C^\infty(\rone)$ functions. In addition, the large $x$ asymptotics for $U,V$ are as follows 
  $$
  U(x)\sim e^{-\sqrt{4-\om^2}|x|}, V(x)\sim e^{-\sqrt{4-9 \om^2}|x|}.
  $$
We now pass to the second order expansion, in terms of $\eps$, for the function $g$. To simplify matters, we work with the additional restriction that the increments $(h_1, h_2)$ satisfy the constraint, 
\begin{equation}
	\label{75} 
	\Re [\dpr{-U''+(4-\om^2)U}{h_1}+\dpr{-V''+(4-\om^2)V}{h_2}]=0.
\end{equation}
As a consequence of this extra property, we have that 
\begin{equation}
	\label{80} 
	I^2[U+\epsilon h_1, V+\epsilon h_2] = 1+ \eps^2(\|h_1'\|^2+\|h'_2\|^2+(4-\om^2) \|h_1\|^2+(4-9\om^2) \|h_2\|^2).
\end{equation}
Next, we find the  second order expansion, in terms of $\eps$, for the functional $J[U+\eps h_1, V+\eps h_2]$. We have 
$$
 J[U+\eps h_1, V+\eps h_2] =  J_0+\eps (A_1[h_1]+A_2[h_2]+A_3[\bar{h}_1]+A_4[\bar{h}_2])+ \eps^2 B[h_1, h_2, \bar{h}_1, \bar{h}_2]+O(\eps^3), 
$$
where $A_1, \ldots, A_4$ are linear forms~\footnote{The particular formulas were actually displayed before, but they become irrelevant here, due to \eqref{80}}, while 
\begin{eqnarray*}
	B[h_1, h_2, \bar{h}_1, \bar{h}_2] &=&  \int\left( 12 U^2 h_1\bar{h}_1  + 3 U^2(h_1^2+\bar{h}_1^2) + 12 V^2 h_2\bar{h}_2  + 3 V^2(h_2^2+\bar{h}_2^2)\right) +\\
&+& \int \left(12 U V (h_1+\bar{h}_1)(h_2+\bar{h}_2)+12 U^2 h_2\bar{h}_2+ 12 V^2 h_1\bar{h}_1\right) +\\
&+& \int \left( 6 U^2 (\bar{h}_1 h_2+\bar{h}_2 h_1)+6 U V (\bar{h}_1^2+h_1^2)\right). 
\end{eqnarray*}
As a consequence of the expansion for $B$ and \eqref{80}, we have that 
\begin{equation}
	\label{90} 
	0\geq g''(0)= 2(B[h_1, h_2, \bar{h}_1, \bar{h}_2]-J_0 (\|h_1'\|^2+\|h'_2\|^2+(4-\om^2) \|h_1\|^2+(4-9\om^2) \|h_2\|^2).
\end{equation}
 It is technically more convenient to pass to the new variables $h_1=p+i q, h_2=r+ i s$, so that we can rewrite \eqref{90} as follows 
 \begin{eqnarray*}
 & & 	\|p'\|^2+\|q'\|^2+\|r'\|^2+\|s'\|^2+(4-\om^2)(\|p\|^2+\|q\|^2)+(4-9 \om^2)(\|r\|^2+\|s\|^2) \geq \\
 	&\geq &  \f{1}{J_0} \int\left( 12 U^2 (p^2+q^2)  +6 U^2(p^2-q^2) +  12 V^2 (r^2+s^2) +6 V^2(r^2-s^2)\right) +\\
 	&+& \f{12}{J_0} \int \left(4 U V p r + U^2 (r^2+s^2)+ V^2(p^2+q^2) \right) + \\
 	&+& \f{12}{J_0} \int \left( U^2 (p r +q s)+ U V (p^2-q^2)\right). 
 \end{eqnarray*}
This last inequality should hold for all increments $(p+i q, r+i s)$ satisfying \eqref{75}, which is equivalent to 
\begin{equation}
	\label{110} 
	\left(\begin{matrix}
		p \\ q \\ r \\s
	\end{matrix}\right)\perp \left(\begin{matrix}
		- U''+(4-\om^2) U \\0 \\ - V''+(4-9 \om^2) V \\ 0
	\end{matrix}\right)
\end{equation}
 The condition $g''(0)\leq 0$ may be expressed  equivalently  as the positivity of the following 
 self-adjoint operator, on the co-dimension one subspace described by \eqref{110}, 
 $$
 \cl:=\left(\begin{matrix}
 	\cl_1& 0 &   - \f{24}{J_0} UV -\f{6}{J_0} U^2 & 0 \\
 	0 & 	\cl_2  &0  & -\f{6}{J_0} U^2\\
 	- \f{24}{J_0} UV  -\f{6}{J_0} U^2 & 0 &\cl_3 & 0\\
 	0 & -\f{6}{J_0} U^2  & 0 & \cl_4
 \end{matrix}\right)
 $$
where 
\begin{eqnarray*}
	\cl_1 &:=& -\p_{xx}+(4-\om^2) - \f{18}{J_0} U^2-\f{12}{J_0} V^2-\f{12}{J_0} UV \\
	\cl_2 &:=& -\p_{xx}+(4-\om^2) - \f{6}{J_0} U^2-\f{12}{J_0} V^2+\f{12}{J_0} UV \\ 
	\cl_3 &:=& -\p_{xx}+(4-9 \om^2) - \f{18}{J_0} V^2-\f{12}{J_0} U^2 \\ 
		\cl_4 &:=& -\p_{xx}+(4-9 \om^2) - \f{6}{J_0} V^2-\f{12}{J_0} U^2 
\end{eqnarray*}

\end{proof}
\noindent Clearly, the assignment 
\begin{equation}
	\label{300} 
\phi:=\f{U}{\sqrt{J_0}},  \psi:=\f{V}{\sqrt{J_0}}
\end{equation}
produces bell-shaped solutions of \eqref{30}, moreover one can infer spectral properties of the corresponding operator, from the properties of the operator $\cl$.  In fact, we have established more, namely that $\dpr{\cl h}{h}\geq 0$, whenever $h\perp \left(\begin{matrix}
	- U''+(4-\om^2) U \\0 \\ - V''+(4-9 \om^2) V \\ 0
\end{matrix}\right)$. Using the rescaling transformation  \eqref{300}, this is equivalent to 
\begin{equation}
	\label{310} 
	\dpr{\ch u}{u}\geq 0, u\perp \left(\begin{matrix}
		- \phi''+(4-\om^2) \phi \\0 \\ - \psi''+(4-9 \om^2) \psi \\ 0
	\end{matrix}\right)
\end{equation}
We state the results in the following corollary. 
\begin{corollary}
	\label{cor:10} 
	The linearized self-adjoint operator operator $\ch$, introduced in \eqref{212}, has exactly one negative eigenvalue, i.e., the property $n(\ch)=1$. 
\end{corollary}

\section{Stability of the waves}
In this section, we consider the eigenvalue problem \eqref{210}. Since the self-adjoint portion of it has the property 
$$
n\left(\begin{matrix}
	\ch & 0 \\ 0 & I 
\end{matrix}\right)=n(\ch)=1,
$$
 according to Corollary \ref{cor:10}, we can apply the formula \eqref{d:30}. To this end, we need to calculate the generalized kernel $gKer( \left(\begin{matrix}
 	0 & -I \\ I & -\cj
 \end{matrix}
\right) \left(\begin{matrix}
 	\ch & 0 \\ 0 & I 
 \end{matrix}\right))$, compute the matrix $\cd$ and then $n(\cd)$. 

It is actually easy to see that for this program, we may as well restrict our attention to the subspace of even functions. Indeed, the eigenvalue problem \eqref{210} splits into two independent eigenvalue problems, posed on the even and odd subspaces. The problem on the odd subspace is in fact a trivial one (i.e., no non-trivial solutions). This can be seen from \eqref{310}, as $\ch$  restricted to the odd subspace is in fact a positive operator, i.e. $\ch|_{X_{odd}}\geq 0$. Indeed, as $\phi, \psi$ are even functions  
$$
X_{odd}\perp \left(\begin{matrix}
	- \phi''+(4-\om^2) \phi \\0 \\ - \psi''+(4-9 \om^2) \psi \\ 0
\end{matrix}\right). 
$$
and hence $\ch|_{X_{odd}}\geq 0$. So, now we proceed to find the adjoint eigenvectors corresponding to the kernel vector $\left(\begin{matrix} 
	0 \\ \phi \\ 0 \\   3 \psi 
\end{matrix} \right) $. We solve 
$$
\left(\begin{matrix}
	0 & -I \\ I & -\cj
\end{matrix}
\right) \left(\begin{matrix}
	\ch & 0 \\ 0 & I 
\end{matrix}\right) \vec{\eta}=\left(\begin{matrix} 
0 \\ \phi \\ 0 \\   3 \psi 
\end{matrix} \right). 
$$
Applying the inverse operator $\left(\begin{matrix}
	-\cj  & I \\ -I & 0
\end{matrix}
\right) $ 
we obtain for $\vec{\eta}=\left(\begin{matrix} \eta_1 \\ \eta_2 \end{matrix} \right)$, 
$$
\eta_2= -\left(\begin{matrix} 
	0 \\ \phi \\ 0 \\   3 \psi 
\end{matrix} \right), \ch \eta_1=-\cj \left(\begin{matrix} 
	0 \\ \phi \\ 0 \\   3 \psi 
\end{matrix} \right)= 2\om  \left(\begin{matrix} 
 \phi \\ 0 \\   9 \psi \\ 0  
\end{matrix} \right). 
$$
Taking derivative with respect to $\om$ in the profile equation \eqref{30} leads to the formula 
\begin{equation}
	\label{320} 
	\ch [\p_\om \left(\begin{matrix} 
		 \phi \\ 0 \\ \psi \\ 0  
	\end{matrix} \right)]= 2\om  \left(\begin{matrix} 
	\phi \\ 0 \\   9 \psi \\ 0  
\end{matrix} \right). 
\end{equation}
Thus, we can take 
$$
\eta_1=\ch^{-1}[2\om  \left(\begin{matrix} 
	\phi \\ 0 \\   9 \psi \\ 0  
\end{matrix} \right)]=\p_\om \left(\begin{matrix} 
\phi \\ 0 \\ \psi \\ 0  
\end{matrix} \right). 
$$
In this fashion, we have identified an element $\vec{\eta}=\left(\begin{matrix} \eta_1 \\ \eta_2 \end{matrix} \right)$ of the generalized kernel, namely
$$
\eta_1= \p_\om \left(\begin{matrix} 
	\phi \\ 0 \\ \psi \\ 0  
\end{matrix} \right), \ \ \eta_2= -\left(\begin{matrix} 
0 \\ \phi \\ 0 \\   3 \psi 
\end{matrix} \right). 
$$
Proceeding to find further elements of $gKer$, we need to solve 
$$
\left(\begin{matrix}
	0 & -I \\ I & -\cj
\end{matrix}
\right) \left(\begin{matrix}
	\ch & 0 \\ 0 & I 
\end{matrix}\right) \vec{z}=\vec{\eta}
$$
Applying again the inverse, we are led to the following system 
$$
\ch z_1=-\cj \eta_1+\eta_2, \ \ z_2=-\eta_1.
$$
This of course requires a solvability condition in the first component, namely 
\begin{equation}
	\label{340} 
	-\cj \eta_1+\eta_2\perp Ker[\ch]=span[\left(\begin{matrix} 
		0 \\ \phi \\ 0 \\   3 \psi 
	\end{matrix} \right)].
\end{equation}
Equivalently,  the second adjoint eigenvector $\vec{z}$ does not exist, whenever the solvability condition \eqref{340} is violated. More precisely, 
\begin{equation}
	\label{350} 
0\neq \dpr{	-\cj \eta_1+\eta_2}{\left(\begin{matrix} 
		0 \\ \phi \\ 0 \\   3 \psi 
	\end{matrix} \right)}=-\left(\om \p_\om\left(\int \phi^2+9 \psi^2\right)+\int \phi^2+9 \psi^2\right)
\end{equation}
Under the condition \eqref{350}, we have that $dim(gKer)=1$, and hence the matrix $D\in\cm_{1\times 1}$, with 
$$
\cd_{11}=\dpr{\left(\begin{matrix}
	\ch & 0 \\ 0 & I 
\end{matrix}\right)\vec{\eta}}{\vec{\eta}}= \dpr{\ch \eta_1}{\eta_1}+\dpr{\eta_2}{\eta_2}=\om\p_\om
\int_{\rone} \left(\phi^2+9 \psi^2\right) dx + \int_{\rone} \left(\phi^2+9 \psi^2\right) dx 
$$
Thus, according to the index counting theory, more precisely \eqref{d:30}, 
the waves are stable in the case of $n(L)=1$, if and only if 
\begin{equation}
	\label{370} 
	\p_\om\left(\om \int_{\rone} \left(\phi^2+9 \psi^2\right) dx \right)=\om \p_\om \int_{\rone} \left(\phi^2+9 \psi^2\right) dx +\int_{\rone} \left(\phi^2+9 \psi^2\right) dx <0
\end{equation}
Equivalently, in this case, the stability is equivalent to the property that 
$$
\om\to \om \int_{\rone} \left(\phi^2+9 \psi^2\right) dx
$$
is a decreasing function. 

In what follows, we will explore the different branches
of the real solutions of Eqs.~(\ref{30}), which can
be used according to the prescription of Eq.~(\ref{20})
in order to develop approximations to the oscillon 
periodic orbits of Eq.~(\ref{10}). We will subsequently
study the stability of these states at the level
of the reduced equations~(\ref{15}) and will then
seek to connect the resulting conclusions with the
original PDE of Eq.~(\ref{10}).

\section{Numerical Results}

\begin{figure}[htb]
\begin{tabular}{cc}
\includegraphics[width=9cm]{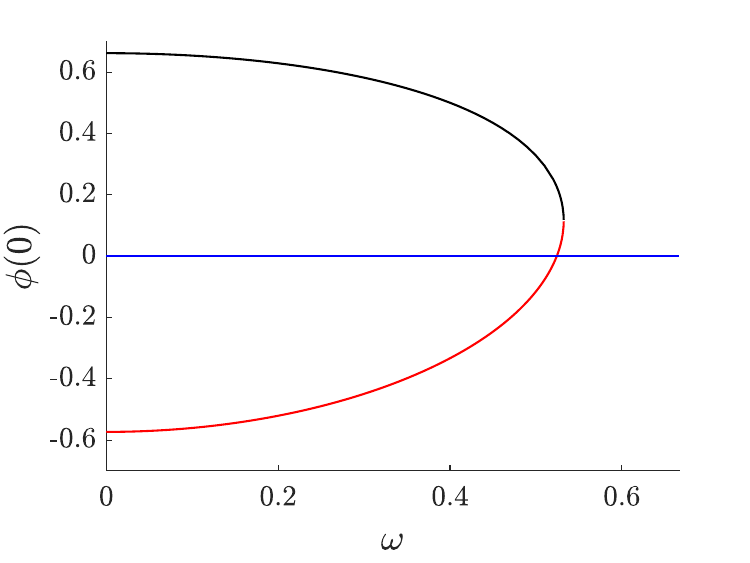} & 
\includegraphics[width=9cm]{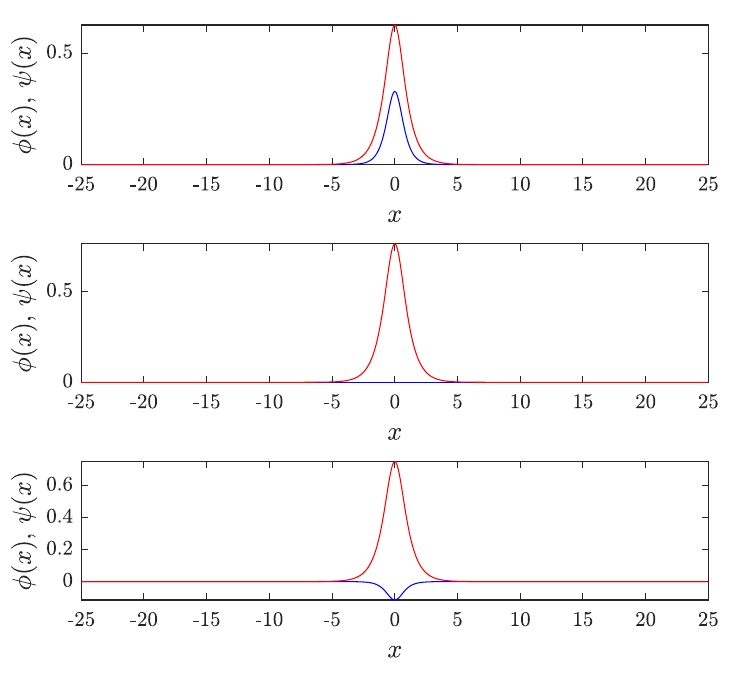}\\
\end{tabular}%
\caption{{(Left panel)} A bifurcation diagram of the solutions to
Eq.~(\ref{30}) that were identified in our analysis.
The value of the first component at $x=0$ ($\phi(0)$) is
used as a bifurcation diagnostic. The blue solid line
represents a $(0,+)$ branch with only the third harmonic
being ``populated''. The black branch is a $(+,+)$ branch
with both components featuring a positive (sign-definite)
waveform. Finally, the red branch is a $(-,+)$ branch
with the two components starting with opposite values,
but the first component crosses $\phi(x)=0$ 
when $\omega\rightarrow\omega_0=0.5242$ leading to both components
becoming positive before this branch collides in a 
saddle-center bifurcation with the other $(+,+)$ branch
at $\omega=\omega_+=0.5323$. {(Right panel) Profiles of the oscillons for the different branches (from top to bottom: $(+,+)$, $(0,+)$ and $(-,+)$ branches) at $\omega=0.5$. Blue (red) curves correspond to the $\phi(x)$ ($\psi(x)$) component.}}
\label{fig:bif}
\end{figure}

In our efforts to identify solutions to Eqs.~(\ref{30}),
we identified 3 branches of possible solutions.
We thus hereafter explore these solution branches that can be 
discerned in the diagram of { the left panel of Fig.~\ref{fig:bif}.  This is complemented by an example of the profile of each of the oscillon branches
that we have identified in the right panel of the figure. 
The former} represents the amplitude of the $\phi(x)$ component,
providing accordingly a useful bifurcation diagnostic. 
The first branch of solutions, depicted in blue, corresponds to 
$(0,+)$ oscillons, i.e., $\phi(x)=0$, $\psi(x)>0$, with 
only the third harmonic being present. These waveforms have
been found to
exist for any $\omega\in(0,2/3)$, which is in line with 
earlier dynamical observations of the original
PDE of Eq.~(\ref{10})~\cite{baras_oscillon}. 
In the latter the frequency of these
waves $3 \omega<2 \Rightarrow \omega<2/3$. 
\footnote{{ It is important to highlight
here that this is the branch analogous to
the solutions presented in~\cite{baras_oscillon},
as there are no solutions of the form $(+,0)$
that are possible in model of Eqs.~(\ref{30}.}}
On the other hand,the red line in Fig.~\ref{fig:bif} corresponds to solutions of the $(-,+)$ form, i.e. $\phi(x)<0$, $\psi(x)>0$, in the interval $\omega\in(0,\omega_0)$ with $\omega_0=0.5242$; as one can see, the amplitude of the $\phi$ component tends to zero when $\omega\rightarrow\omega_0$ and becomes zero just at $\omega=\omega_0$. For $\omega>\omega_0$, $\phi(x)>0$ {\it
and} $\psi(x)>0$, meaning that the solution, in our
notation, becomes of the $(+,+)$ form,  until the
solution disappears for only slightly larger 
frequency, i.e., at $\omega=\omega_+=0.5323$ through a saddle-center bifurcation with a third branch, also of the $(+,+)$ form and depicted in black. In particular, the branch depicted with
a black line has a larger (positive) amplitude within its $\phi$ component,
while the red one has a lower amplitude and the two 
collide and disappear hand-in-hand at $\omega=\omega_+=0.5323$. However,  this saddle-center
event
is not the only bifurcation occurring. As the
$(-,+)$ branch turns to a $(+,+)$ one with 
$\phi(x)$ crossing through 0, this red branch
``collides'' in a transcritical-like bifurcation with symmetry,
as we will see below, with the blue branch of the
form $(0,+)$. We will explore the implications
of these bifurcations for the branches involved
in what follows.

\begin{figure}[htb]
\begin{tabular}{cc}
\includegraphics[width=9cm]{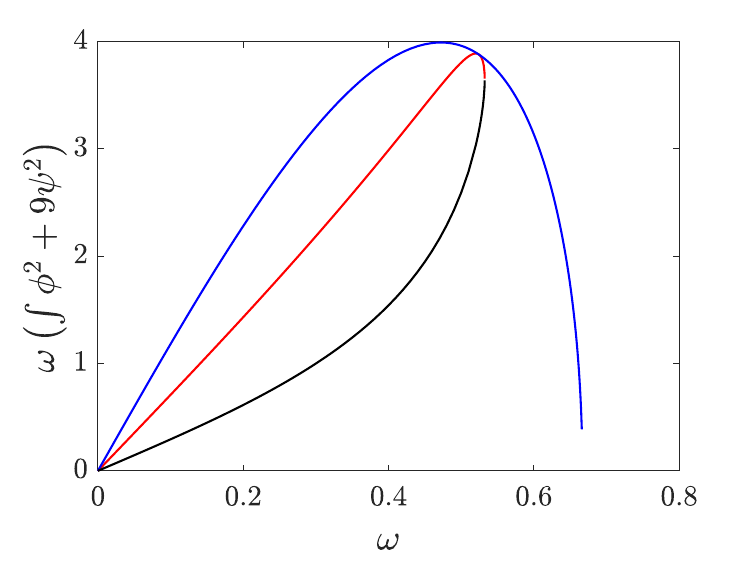} &
\includegraphics[width=9cm]{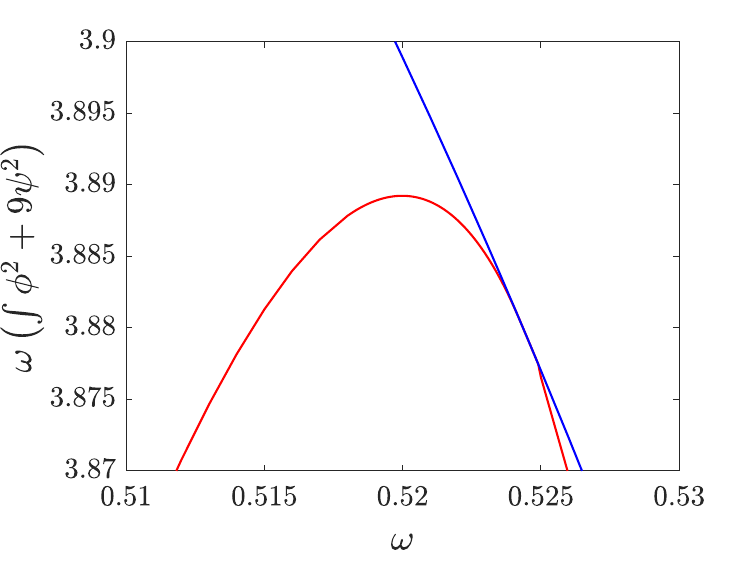} \\
\end{tabular}%
\caption{In this figure, we demonstrate the
$\omega \int_{\rone} \left(\phi^2+9 \psi^2\right) dx$ quantity that
we associated in Theorem~\ref{theo:20} with stability.
The red $(-,+)$ branch changes monotonicity (and hence 
stability) at $\omega=0.52$; see also the zoom-in
of the right panel. The black $(+,+)$ branch is always
increasing and hence always bears real eigenvalues.
The blue $(0,+)$ branch has a change of monotonicity
at $\omega=0.4719$ analyzed further below.}
\label{fig:VK}
\end{figure}

In Fig.~\ref{fig:VK} we show the dependence with $\omega$ of the VK-like quantity discussed earlier,
namely $\omega \int_{\rone} \left(\phi^2+9 \psi^2\right)  dx$. We observe a maximum of the $(0,+)$ oscillon
of the blue branch at $\omega=0.4719$ and another one for the $(-,+)$ oscillon of the red branch at $\omega=0.5200$. 
Interestingly, for the blue branch, we will see that
the change of monotonicity of the relevant quantity
is associated with  a
real eigenvalue changing to imaginary, as the
relevant quantity changes from increasing to decreasing.
On the other hand, for the black $(+,+)$ branch, the 
``wrong'' monotonicity (i.e., its increasing nature)
implies that this branch will
always feature real eigenvalue pairs, as we will see
to be the case in what follows. Interestingly,
the situation for the red branch will be seen in
our eigenvalue analysis to be more complicated. 

\begin{figure}[htb]
\begin{tabular}{cc}
\includegraphics[width=9cm]{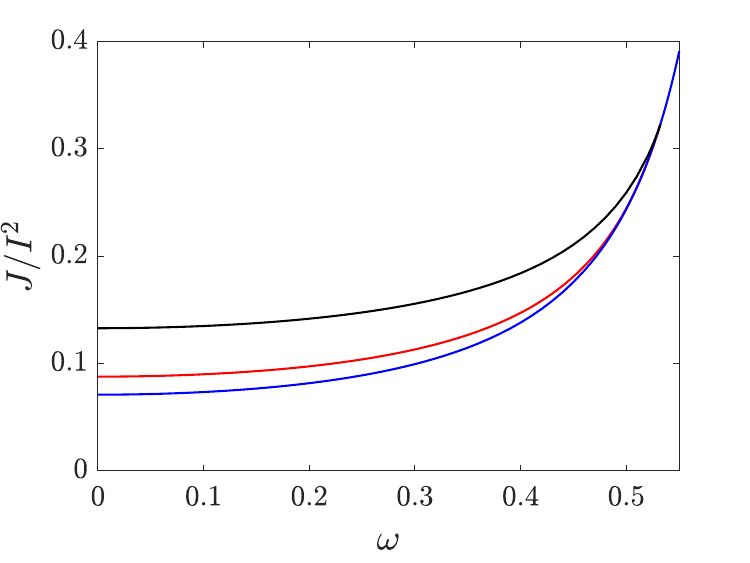} &
\includegraphics[width=9cm]{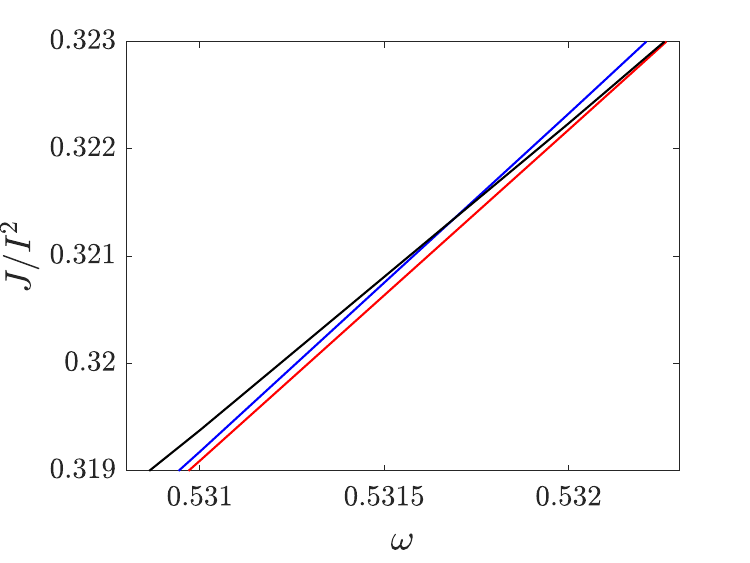} \\
\end{tabular}%
\caption{The quantity $J/I^2$ where $J$ and $I$ 
are defined in Proposition~\ref{prop:10} is given
on the left panel for each of the previously discussed
branches. The right panel is again a zoom-in where
a crossing of this quantity for the black and blue
branch takes place at $\omega<0.5316$.}
\label{fig:maximizer}
\end{figure}

Another quantity that was deemed central to our
theoretical analysis of Section~\ref{construct},
when constructing the waves was the functional
$J/I^2$ where $J$ and $I$ are defined in
Proposition~\ref{prop:10}. 
Accordingly, in Fig.~\ref{fig:maximizer}, we
illustrate the quantity $J/I^2$ versus $\omega$. One can see that the largest value of the relevant quantity
pertains to the black $(+,+)$ branch for $\omega<0.5316$,
when the blue branch acquires the largest
$J/I^2$ value and becomes the global maximizer until
the branch ceases to exist at $\omega=2/3$.
The relevant global maximizer is unstable throughout
its existence (irrespectively of whether it pertains
to the black or blue branch), as we will confirm
in our eigenvalue analysis.

\begin{figure}[htb]
\begin{tabular}{c}
\includegraphics[width=9cm]{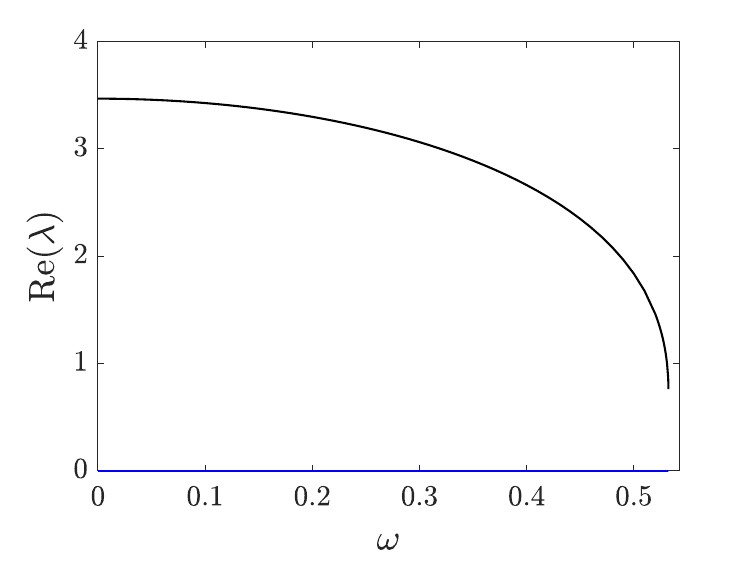} \\
\end{tabular}%
\caption{Dependence of the real part of the eigenvalues of the $(+,+)$ family with respect to the frequency. The black line indicates that such a real eigenvalue exists for
all the frequencies for which the branch exists, thus 
rendering it spectrally unstable.}
\label{fig:stab++}
\end{figure}

\begin{figure}[htb]
\begin{tabular}{cc}
\includegraphics[width=9cm]{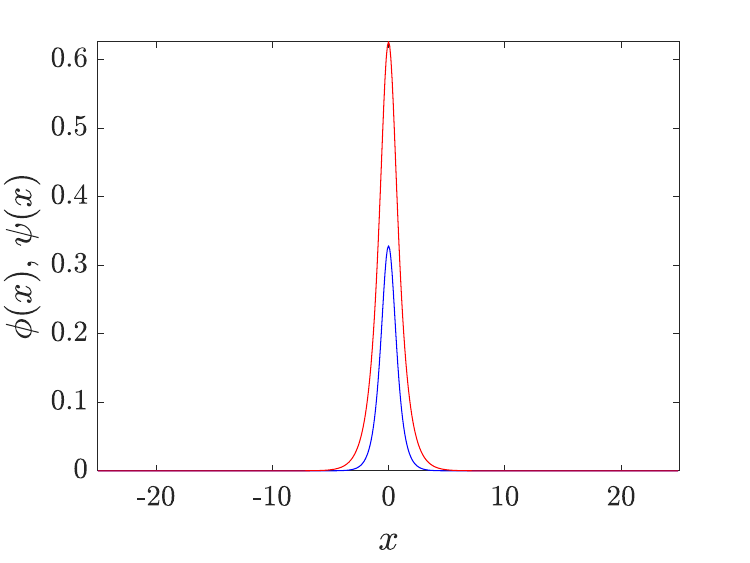} &
\includegraphics[width=9cm]{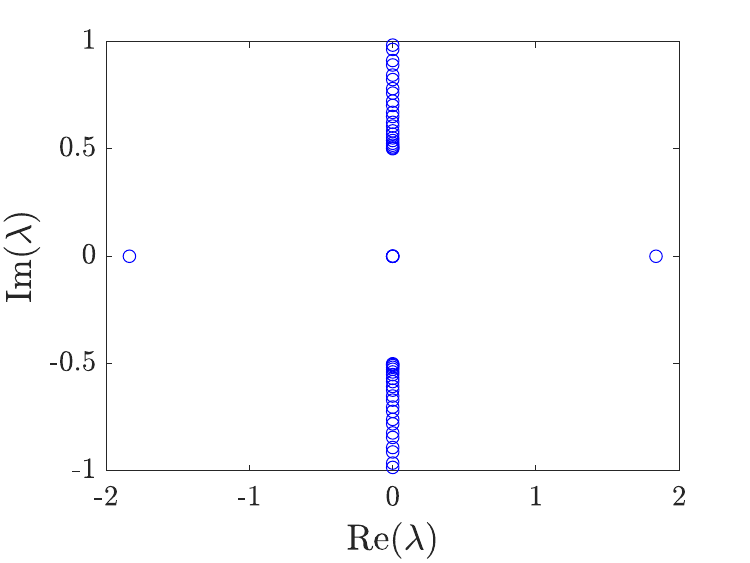} \\
\end{tabular}%
\caption{Oscillon in the $(+,+)$ branch with $\omega=0.5$
(left panel) and the corresponding spectral plane
featuring a real (unstable) eigenvalue.}
\label{fig:examp1}
\end{figure}

We now turn to an analysis of the eigenvalues of
the different branches to help elucidate their stability
and connect them to the VK-criterion discussed 
in Theorem~\ref{theo:20}. As discussed already above,
and in line with the VK criterion of the Theorem,
the $(+,+)$ branch is always unstable because of a real eigenvalue pair. Fig.~\ref{fig:stab++} shows the 
relevant real part of the (maximal) eigenvalue of
the linearization around such solutions. 
A typical example of the relevant solution and
its spectral place is shown in Fig.~\ref{fig:examp1}.

\begin{figure}[htb]
\begin{tabular}{cc}
\includegraphics[width=9cm]{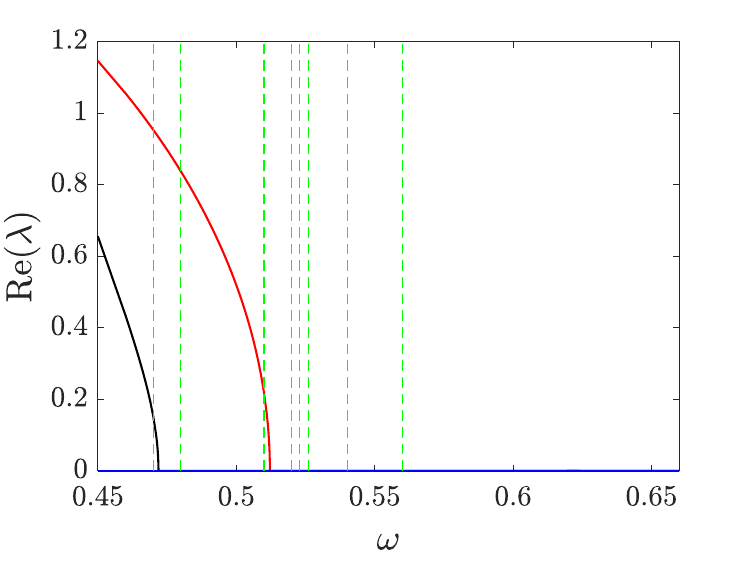} &
\includegraphics[width=9cm]{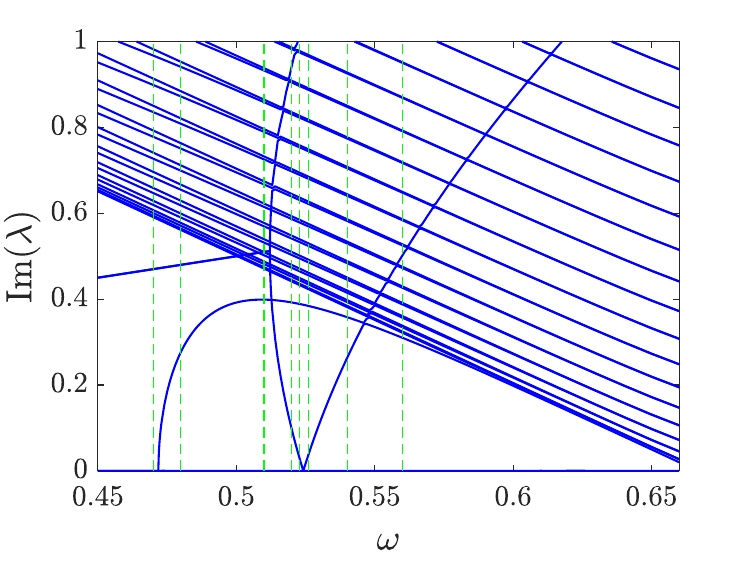} \\
\end{tabular}%
\caption{Dependence of the real (left panel) and imaginary (right panel) parts of the eigenvalues of the $(0,+)$ family with respect to the frequency $\omega$. A black line indicates that an eigenvalue is real and red lines are
associated with eigenvalues with non-zero imaginary part,
i.e., complex ones. Vertical lines correspond to the values of $\omega$ for which the spectral plane is represented in Fig.~\ref{fig:examp2}. Notice that the solution is
fully spectrally stable for $\omega>0.5121$.}
\label{fig:stab0+}
\end{figure}

From the representation of the real and imaginary parts of the spectrum for the $(0,+)$ solutions in Fig.~\ref{fig:stab0+}, one can observe that such a solution is unstable for $\omega<0.5121$, as at such a value of the frequency parameter, there is a Hopf bifurcation where a quartet of complex eigenvalues becomes imaginary, while at $\omega=0.4719$, 
a real eigenvalue pair becomes imaginary; in addition, at $\omega=\omega_0 \equiv 0.5242$ there is an eigenvalue that becomes zero although it is imaginary past this value. 
This is fairly remarkable as what happens here is that
the positive and negative imaginary parts of the eigenvalue 
``exchange sides'' (i.e., the positive becomes negative
and vice versa) {\it without} the state changing its 
stability at the transcritical-like point $\omega_0$.
We will use here the term ``transcritical-like'', since 
in a regular transcritical bifurcation, the two 
branches exist both before and after their collision
(where they become identical), yet they exchange stability
at the critical point. Here, the
branches involved collide (meaning that they exist
before and after the critical point and coincide at 
that point), yet due to each of them
involving a pair of eigenvalues, their stability is
not exchanged; rather the relevant eigenvalue
pairs switch sides (positives become negatives and
vice versa, for real eigenvalues in one of the branches
and imaginary ones in the other).

\begin{figure}[htb]
\begin{tabular}{cc}
\includegraphics[width=7cm]{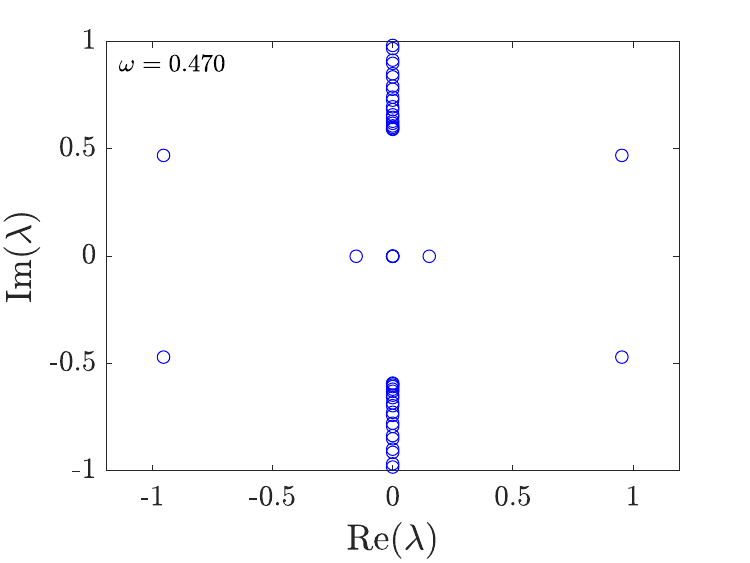} &
\includegraphics[width=7cm]{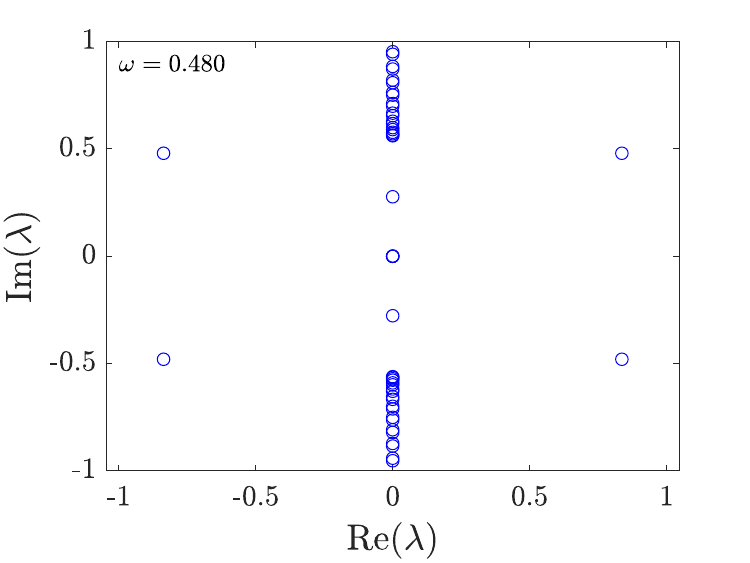} \\
\includegraphics[width=7cm]{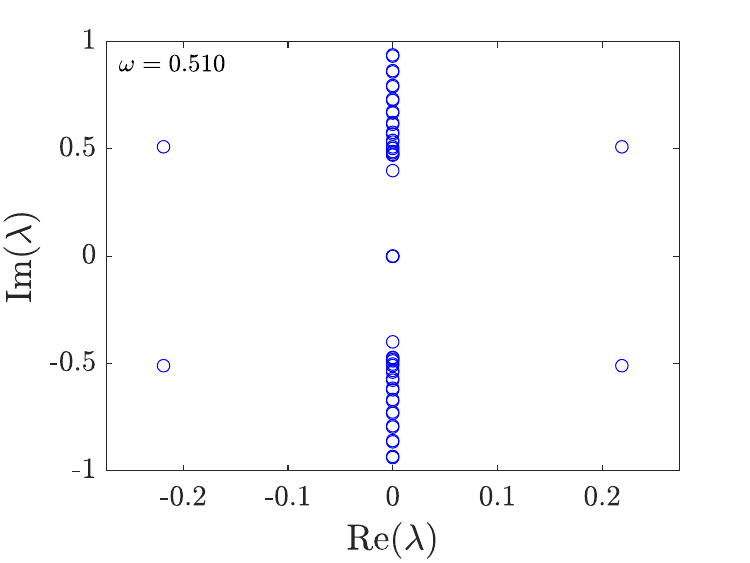} &
\includegraphics[width=7cm]{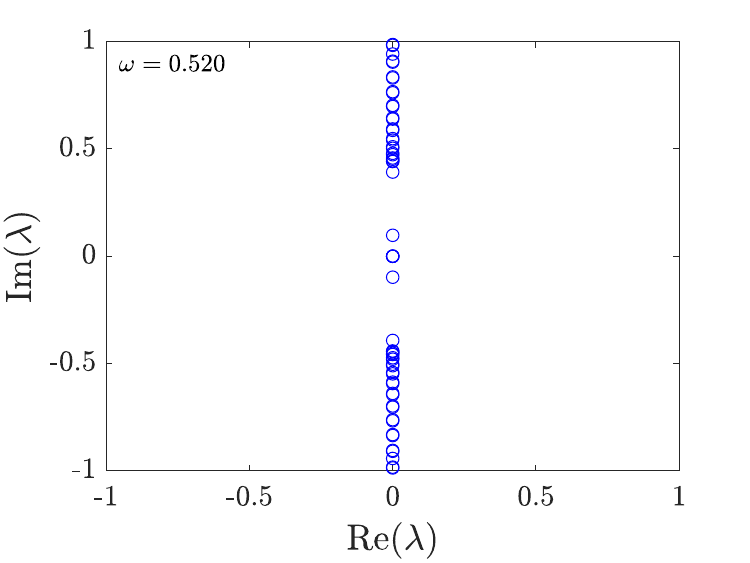} \\
\includegraphics[width=7cm]{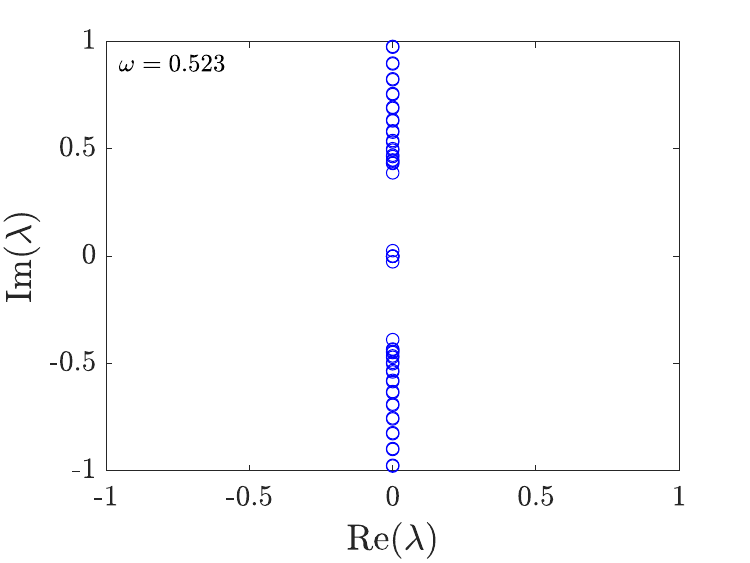} &
\includegraphics[width=7cm]{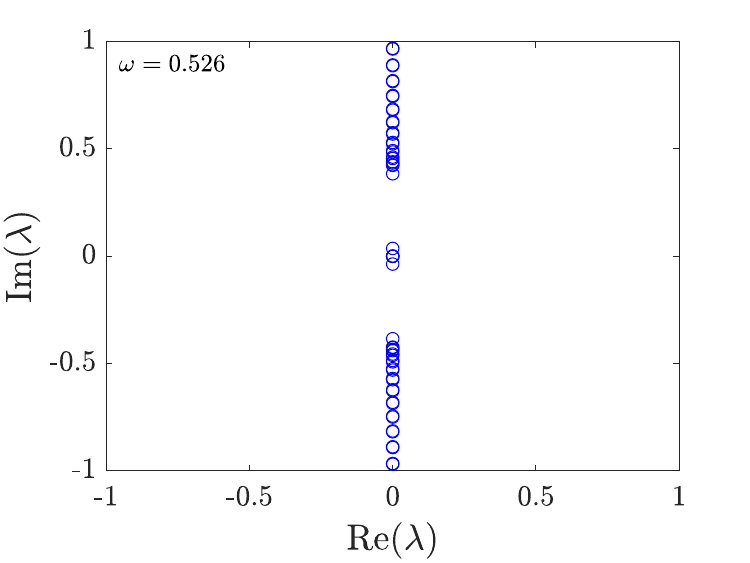} \\
\includegraphics[width=7cm]{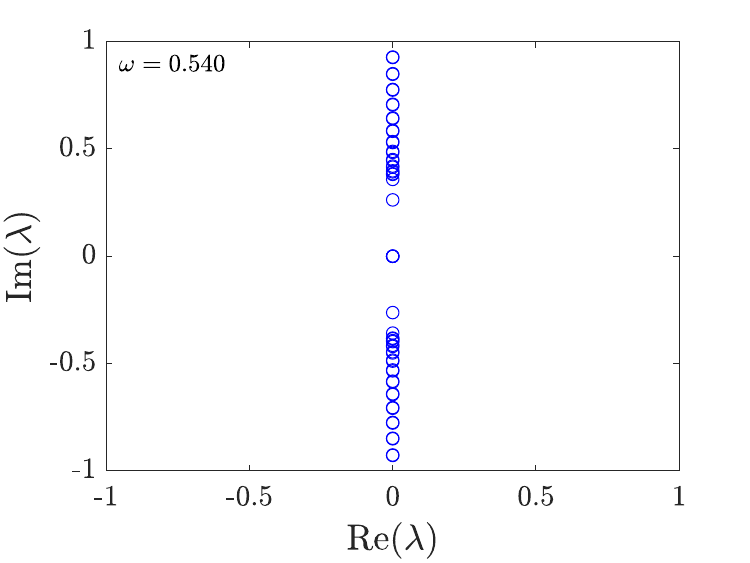} &
\includegraphics[width=7cm]{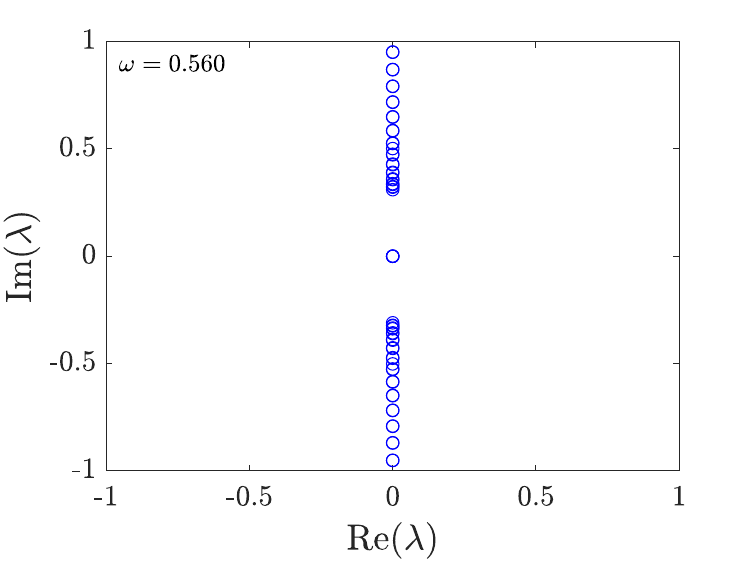} \\
\end{tabular}%
\caption{Oscillons in the $(0,+)$ branch whose frequency corresponds to the green lines in Fig.~\ref{fig:stab0+}. Each value of the frequency is displayed at the corresponding panel.}
\label{fig:examp2}
\end{figure}

Importantly, it is relevant to note that these
results are semi-quantitatively in line with the findings of~\cite{baras_oscillon} at the level of the
full original PDE of Eq.~(\ref{10}), as the latter
findings suggest that the single frequency waveform
is dynamically robust for $\sqrt{2}/3 \equiv 0.4714<
\omega<2/3$ in our notation. We indeed identify
an eigenvalue transition at $0.4719$ (from real
to imaginary) very close to the above threshold,
while the complete stabilization of this steady
state (when it becomes devoid of Hamiltonian Hopf
bifurcations) occurs at $\omega=0.5121$.
Some typical examples of the corresponding
the $(0,+)$ solution branch spectral planes 
are shown in Fig.~\ref{fig:examp2}.

\begin{figure}[htb]
\centering
\begin{tabular}{cc}
\includegraphics[width=0.48\textwidth]{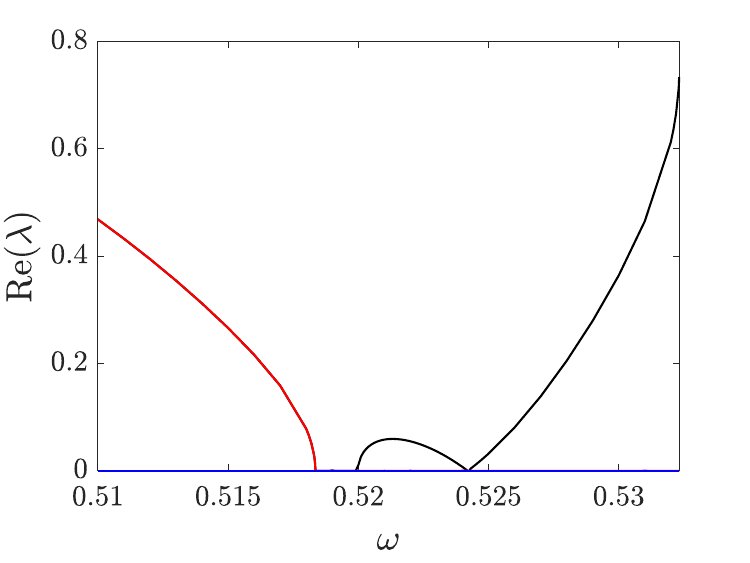} &
\includegraphics[width=0.48\textwidth]{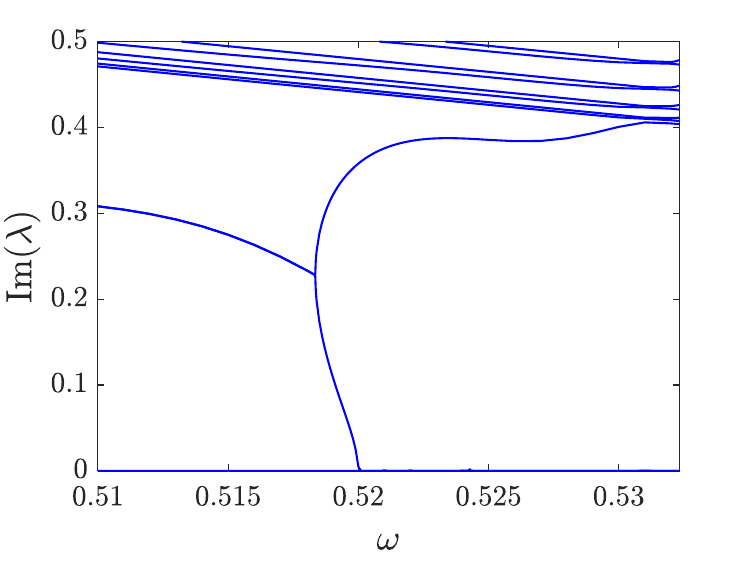} \\
\end{tabular}
\caption{Dependence of the real (left panel) and imaginary (right panel) parts of the eigenvalues of the $(-,+)$ family on the frequency. The black line indicates a real eigenvalue, while the red lines correspond to eigenvalues with nonzero imaginary part.}
\label{figstabmp}
\end{figure}

Finally, as can be deduced from Fig.~\ref{figstabmp}, the $(-,+)$ solutions are unstable for $\omega<0.5183$ as at such a value, the oscillon becomes stable via a Hopf bifurcation; i.e., a quartet turns into two imaginary
pairs at this frequency.
The relevant branch is stable for a very narrow
frequency interval until $\omega=0.5200$ where 
an eigenvalue pair turns from imaginary to real,
leading to the subsequent destabilization of the
branch. 
Notice that we have checked the latter to be
consonant with the Hamiltonian-Krein index theory, given
that for this branch $n({\cal H})=2$, while upon change
of monotonicity $n({\cal D})=1$, hence per the relevant
index theory~\cite{KapitulaPromislow2013}, there should
exist a real eigenvalue pair, which is what we observe.

As $\omega$ increases further, again
very proximally in frequency, i.e., at 
$\omega=\omega_0=0.5242$, the  collision of the $(-,+)$ and
the $(0,+)$ branch discussed above takes place. However, importantly,
this is a rather special example of a 
transcritical-like
bifurcation {\it with symmetry}. That is to say, as 
we pointed out above, 
 for the $(-,+)$ branch, the eigenvalues of the real pair
{exchange sides while both remain real.}
Accordingly, the $(-,+)$  branch before and after the bifurcation
remains unstable, 
just like the $(0,+)$ branch with which it collides
preserves its stability before and after the bifurcation.
At the critical point for the $(-,+)$, the $\phi(x)$
component crosses through zero (from negative to positive,
as the frequency increases). 

\begin{figure*}[htb]
\includegraphics[width=0.48\textwidth]{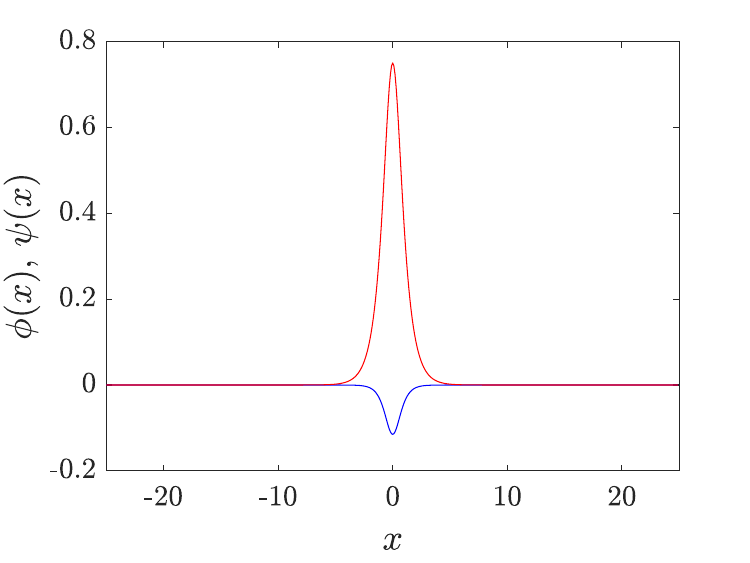}%
\hfill
\includegraphics[width=0.48\textwidth]{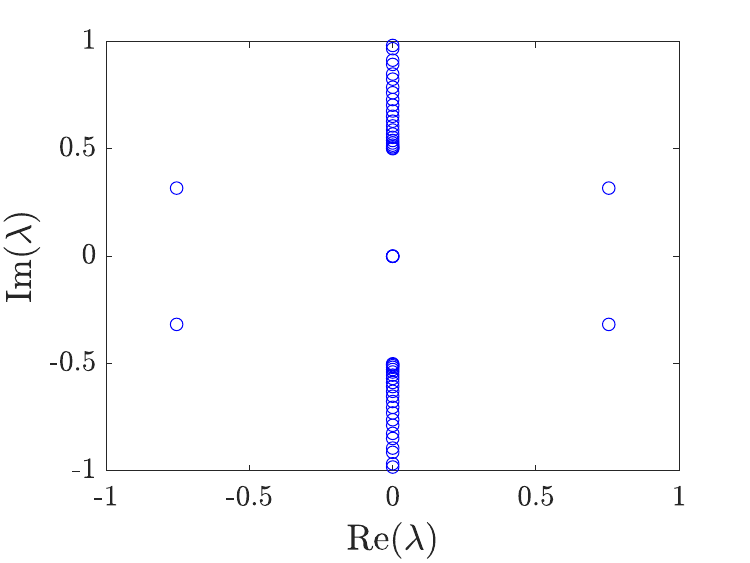}

\includegraphics[width=0.48\textwidth]{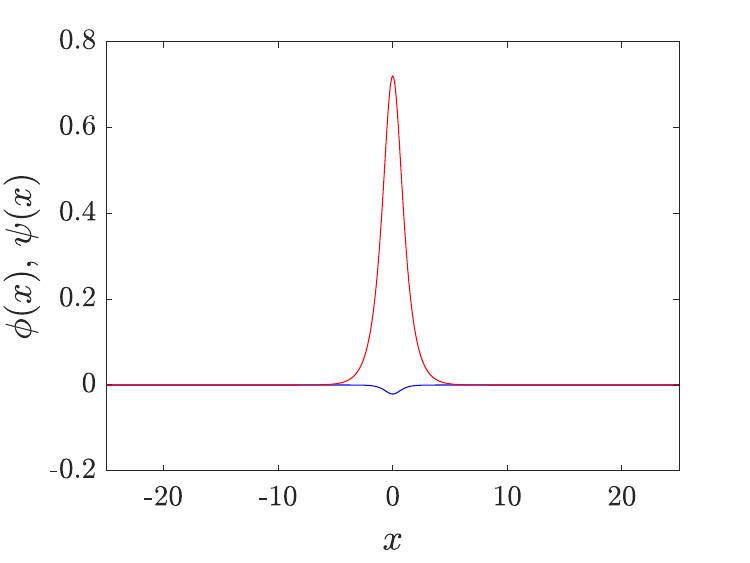}%
\hfill
\includegraphics[width=0.48\textwidth]{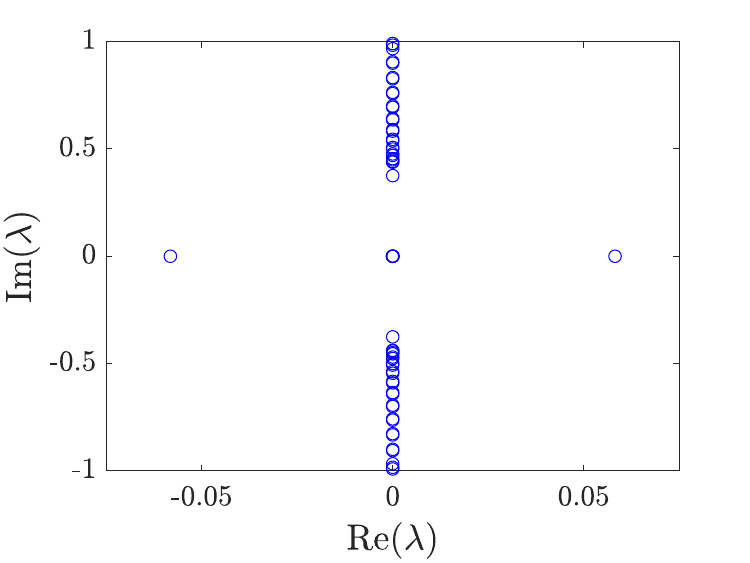}

\includegraphics[width=0.48\textwidth]{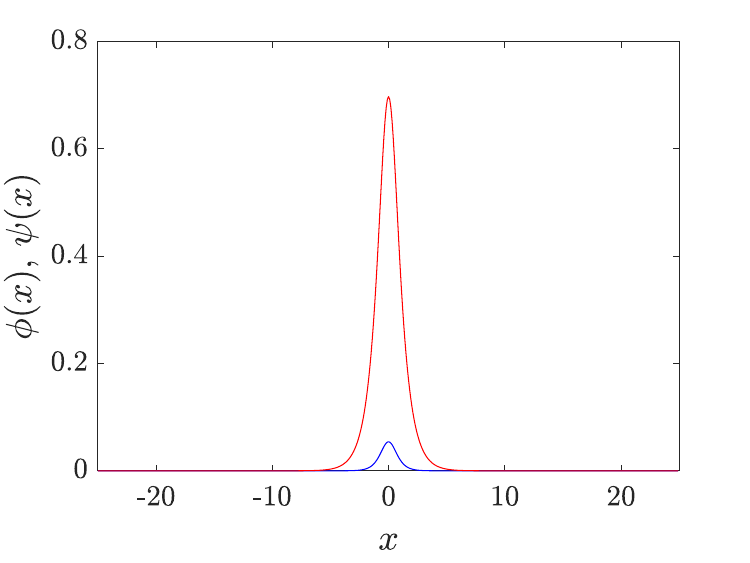}%
\hfill
\includegraphics[width=0.48\textwidth]{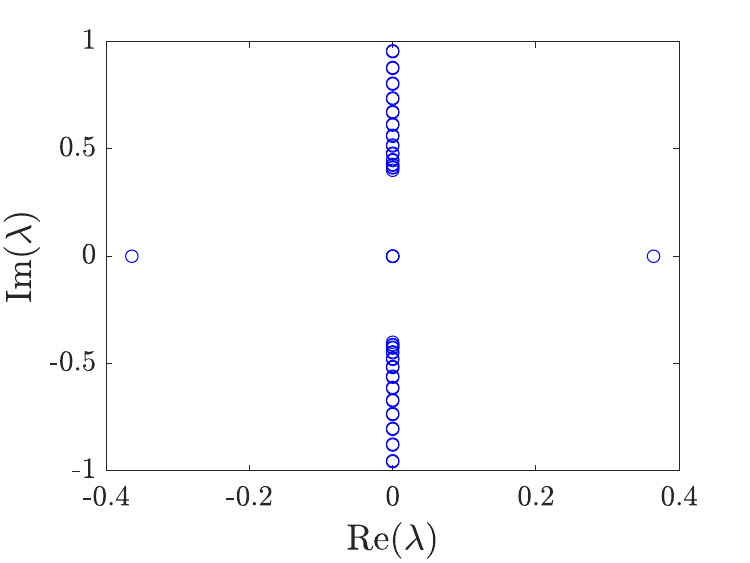}

\caption{Oscillons in the $(-,+)$ branch with $\omega=0.5$ (top panels), $\omega=0.521$ (middle panels), and $\omega=0.53$ (bottom panels). The left panels show the solution profile, while the right ones show the corresponding spectral plane associated with the branch's stability.}
\label{fig:examp3}
\end{figure*}

The oscillons obtained through the solution of 
Eqs.~(\ref{30}) can be used as seeds for fixed point methods that allows to solve the original KG PDE of Eq.~(\ref{10}) in the form of periodic orbits (breathers) of the form

\begin{equation}
    z(t,x)=2\sum_{k=1}^\infty y_k(x)\cos\left[(2k-1)\omega_b t\right]
    \label{psik}
\end{equation}

The main drawback of working with breathers is that they do not exist in a genuinely infinite domain because of resonances of $k>1$ harmonics with the linear modes band. However, in a finite domain, there can appear gaps in the spectrum that can be 
controlled by the size of the domain, whose inverse controls
the discretization in wavenumbers (and accordingly in frequencies). Then, breathers whose harmonics lie in the
resulting frequency gaps have a short interval of stability that is dependent of the domain size and the discretization parameter. In addition, the only oscillon family which gives rise to breathers with features similar to that of the original oscillon is that
of the $(0,+)$ solutions above. 
Because of this, we have only considered breathers coming from the $(0,+)$ branch; in that case, the seed for the fixed point algorithm (Newton-Raphson) is taken as $y_1(x)=\psi(x)$ and $y_k(x)=0$ for $k\geq2$, and $\omega_b=3\omega$. In this way, one can find breathers that can potentially be stable as we now discuss.

\begin{figure}[htb]
\begin{tabular}{cc}
\includegraphics[width=9cm]{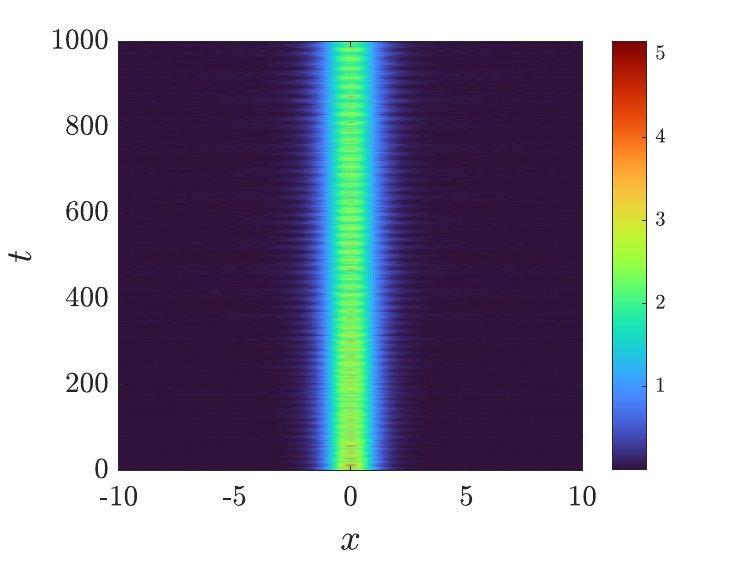} &
\includegraphics[width=9cm]{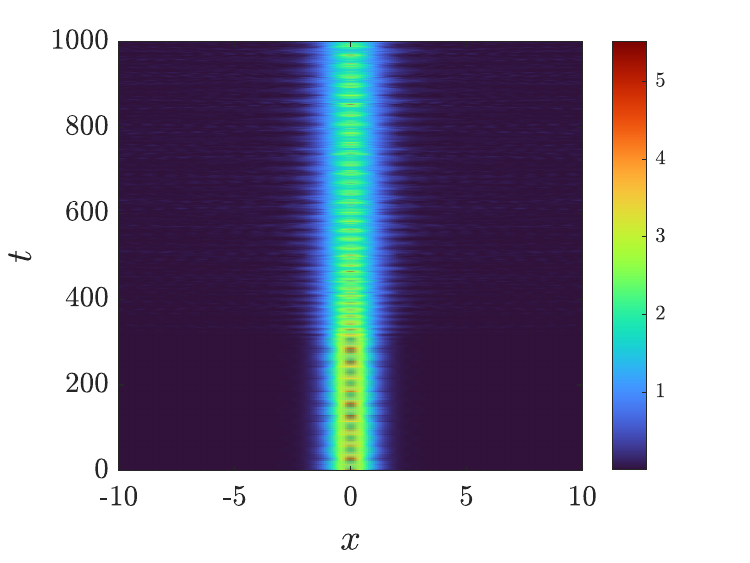} \\
\includegraphics[width=9cm]{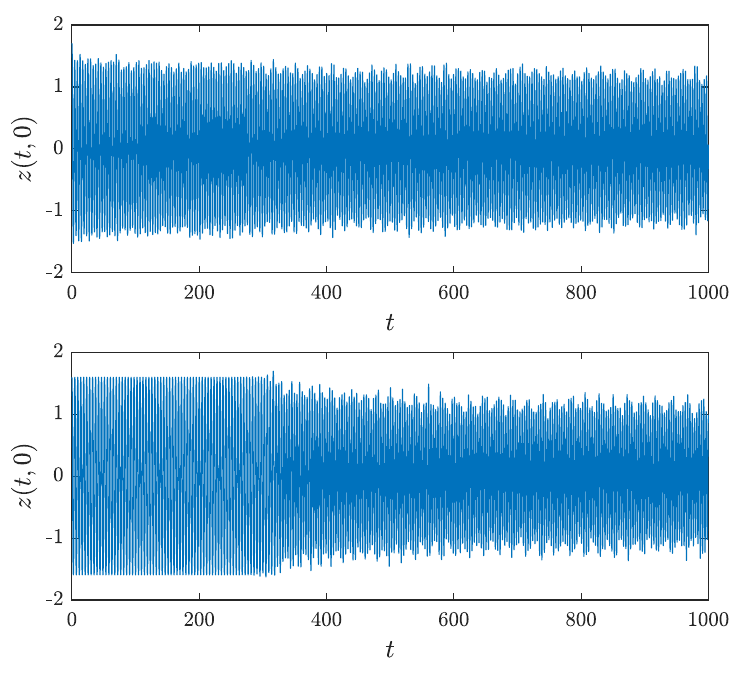} &
\includegraphics[width=9cm]{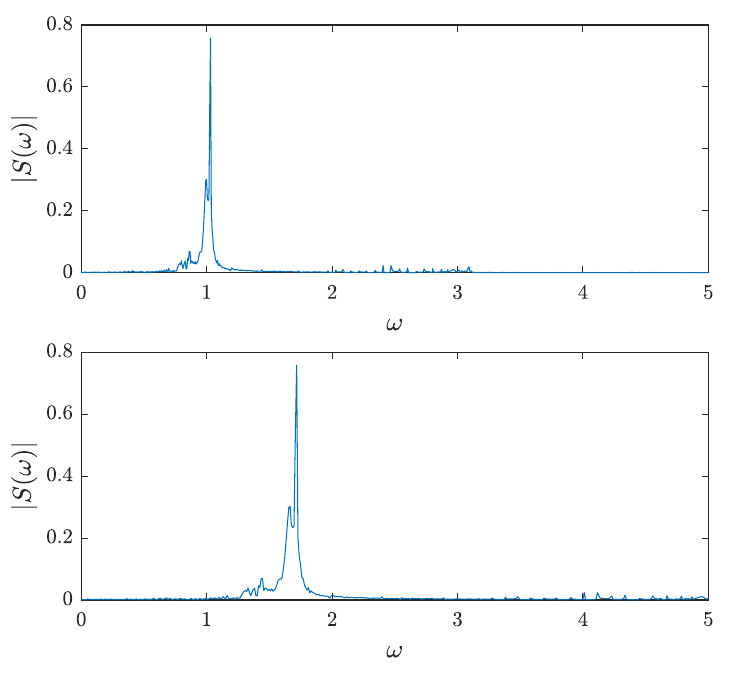} \\
\multicolumn{2}{c}{
\includegraphics[width=9cm]{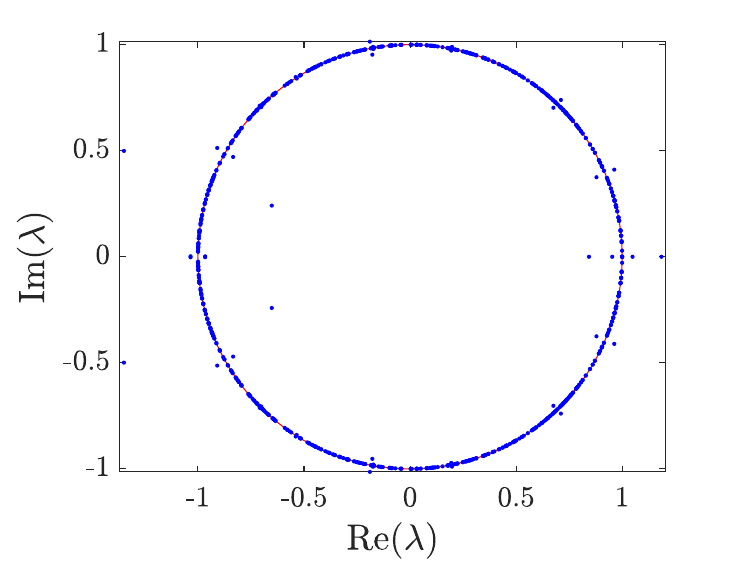}} \\
\end{tabular}%
\caption{(Top panels) Spatio-temporal evolution of the energy density for the KG equation using as initial condition a $(0,+)$ oscillon with frequency $\omega=0.45$ (left) and a (genuinely time-periodic) breather with frequency $3\omega$ (right). The middle panels show the time evolution of the central site and also its Fourier spectrum; top subpanels correspond to the oscillon, and bottom ones to the breather. The bottom panel displays the Floquet spectrum of the breather.}
\label{fig:simul1}
\end{figure}

Based on the resulting solutions, we have considered the dynamics of the KG PDE by taking as initial condition $z(0,x)$ a breather with frequency $3\omega$, i.e., $z(0,x)=2\sum_k y_k(x)$ and comparing it with that stemming from an oscillon, i.e. $z(0,x)=2(\phi(x)+\psi(x))$. 
{ It is worthwhile to note
that in all the relevant examples where the
KG PDE is considered, per the form of the
obtained breathers, the velocity initial
condition $z_t(0,x)$ is set to zero; cf. also
Eq.~(\ref{psik}).}
When considering the $(0,+)$ solution, naturally, $\phi(x)=0$.
Fig.~\ref{fig:simul1} considers the dynamics when taking as initial condition an oscillon with $\omega=0.45$ and comparing
it with the evolution of a breather with frequency $3\omega$; to this aim, a spatio-temporal diagram is shown together with the spectrum of the Floquet operator of the breather, from which one can see that the dominant instability is of exponential nature. The evolution of the central site and its Fourier spectrum shown also complements the picture that can be summarized as the emergence of a quasi-periodic breather-like structure that in the case of the initial breather is created after a transient. From the Fourier spectra one can see that the dominant frequency for the dynamics emerging from the oscillon (breather) is $1.031$ ($1.717$) and a secondary peak of the spectrum is observed around $0.999$ ($1.665$). Although the dynamics draws a number of
parallels between the two cases, there are also some notable
differences. In the approximate oscillon case generated
by the steady state solution of~\ref{30}, we observe
a continuously, yet very slowly decaying structure.
On the other hand, the obtained numerically exact solution
is indeed a periodic orbit, up to numerical tolerance error, yet this error
suffices to eventually destabilize the originally genuinely periodic
evolution for sufficiently long times. The resulting structures
are similar in nature, albeit with different frequencies
involved in the long term dynamics.

\begin{figure}[htb]
\begin{tabular}{cc}
\includegraphics[width=9cm]{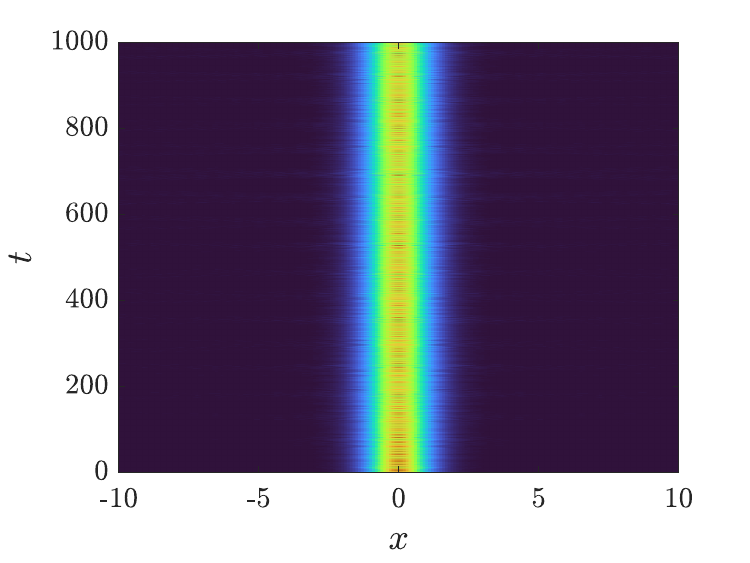} &
\includegraphics[width=9cm]{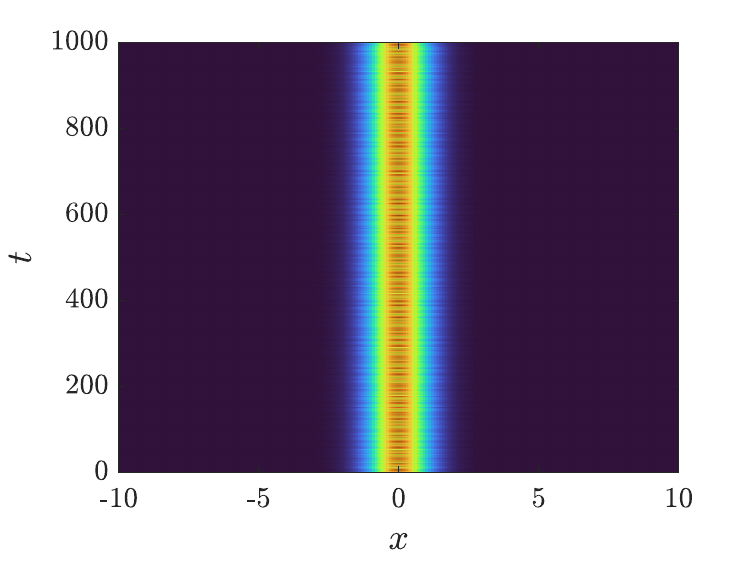} \\
\multicolumn{2}{c}{
\includegraphics[width=9cm]{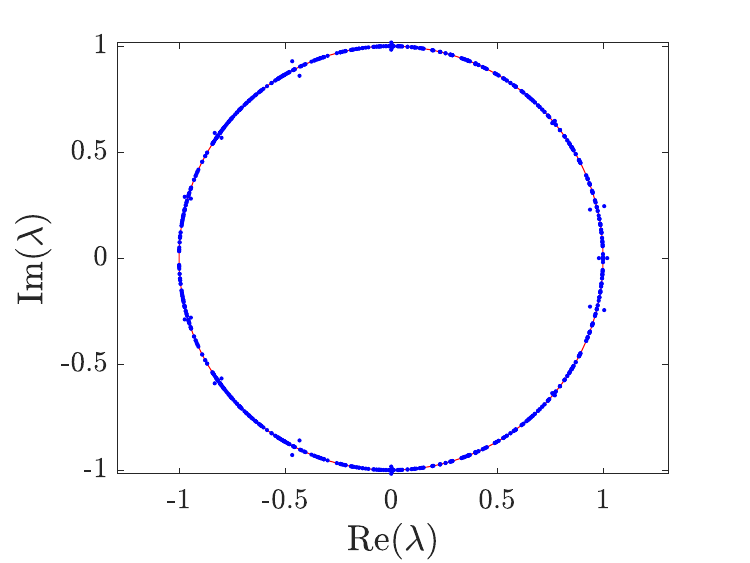}} \\
\end{tabular}%
\caption{Spatio-temporal evolution of the energy density for the KG equation using as initial condition a $(0,+)$ oscillon with frequency $\omega=0.5$ (left panel) and a breather with frequency $3\omega$. The bottom panel displays the Floquet spectrum of the breather.}
\label{fig:simul2}
\end{figure}

\begin{figure}[htb]
\begin{tabular}{cc}
\includegraphics[width=9cm]{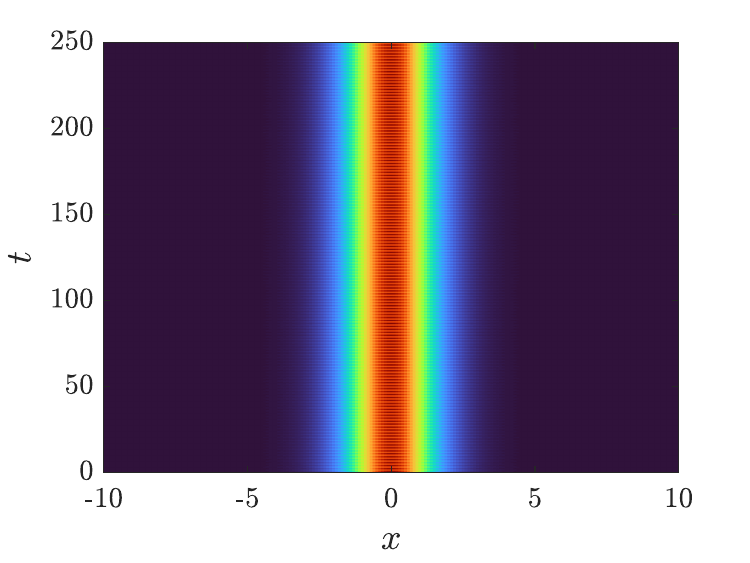} &
\includegraphics[width=9cm]{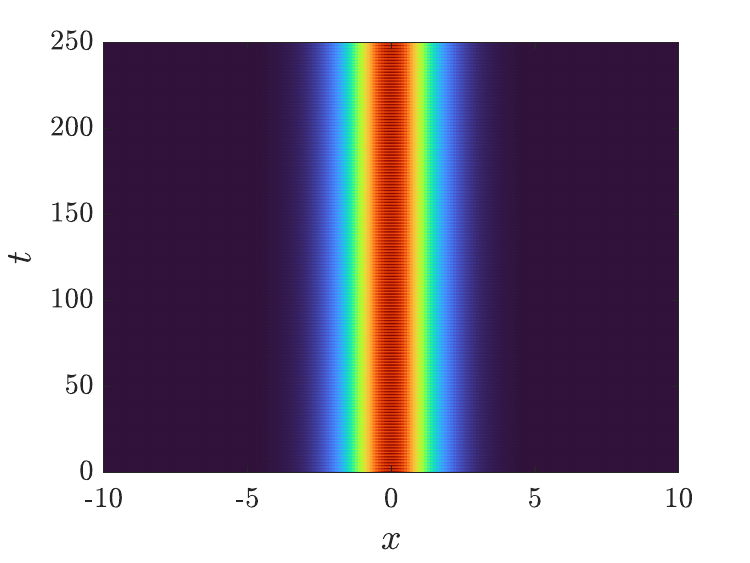} \\ %
\includegraphics[width=9cm]{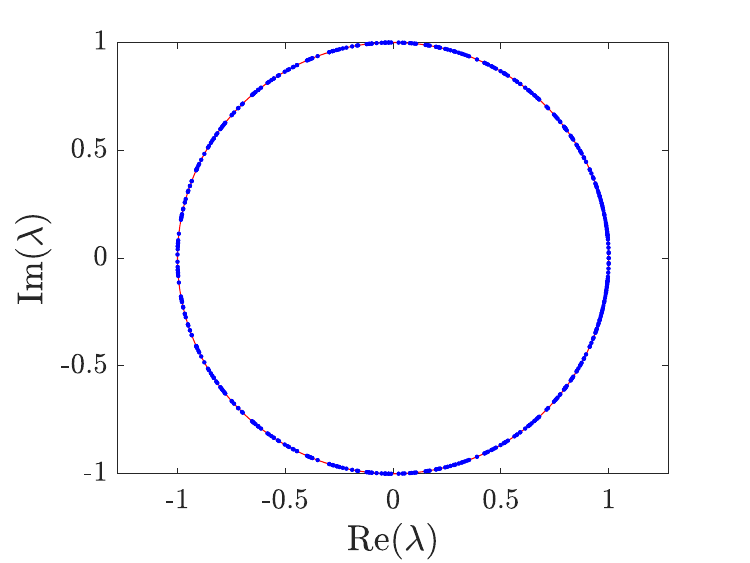} &
\includegraphics[width=9cm]{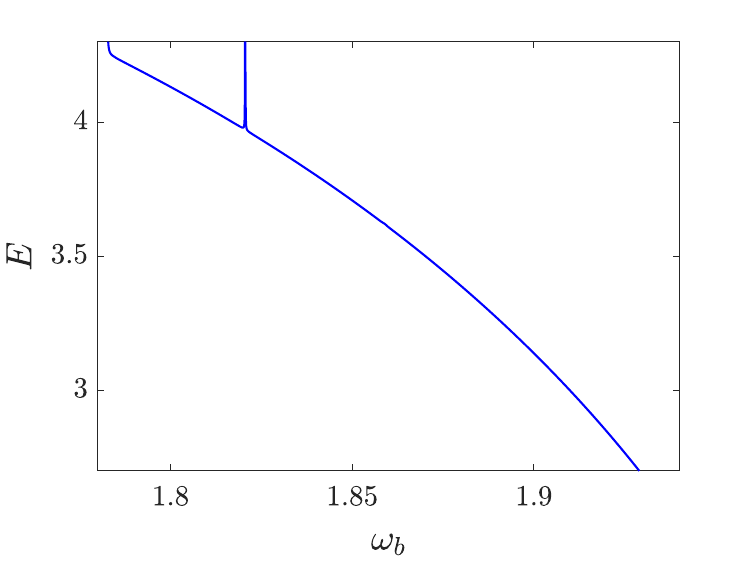} \\
\end{tabular}%
\caption{Spatio-temporal evolution of the energy density for the KG equation using as initial condition a $(0,+)$ oscillon with frequency $\omega=0.61$ (left panel) and a breather with frequency $\omega_b=3\omega=1.83$.  The bottom left panel displays the Floquet spectrum of the breather and the bottom right panel shows the dependence of the energy versus the breather frequency 
for values of $\omega_b$ near $1.83$.}
\label{fig:simul3}
\end{figure}

Figure~\ref{fig:simul2} shows the outcome for the oscillon with $\omega=0.5$ and the breather counterpart with frequency $3\omega$. In this case, the instabilities of the breather are very weak and its shape remains almost invariant for the long time
evolutions of the order of 1000 time units reported. 
It is nevertheless relevant to point out that in this case
too, the genuine breather solution is expected to be 
destabilized for sufficiently long times.
Finally, in Fig.~\ref{fig:simul3} one can observe that for $\omega=0.61$, both the oscillon and the breather are stable, and the oscillon dynamics is quite similar to that of the breather. In addition, we observe the stability of the breather from its Floquet spectrum; in fact, in the bottom right panel of Fig.~\ref{fig:simul3} one can see the dependence of the energy versus the breather frequency, and observe the existence of the typical ladders (``spikes'') caused by the resonance with phonons,
i.e., with linear continuous spectrum modes. 
{ It is relevant to note here that in the context
of oscillons, such a ladder structure has also
been reported in the realm of a three-dimensional
radial oscillon in a ball in the study of~\cite{igor3}.}
Interestingly, for $\omega_b\gtrsim1.82$, the resonances are so small that
they cannot be discerned and the breather is stable for that 
branch. Prior to the resonance at $\omega_b=1.8199$, the breather
solution is stable in the decreasing portion of the branch,
while it is unstable in the (narrow) increasing portion 
evident in Fig.~\ref{fig:simul3}. While the presence of
the continuous spectrum resonances renders  the
continuation of the relevant breather branch for all
frequencies tricky, our expectation is that for $3 \omega>\sqrt{2}$,
the breather branch will have a character similar to what
is reported above with a typical marginally stable Floquet
spectrum (in the exception of small finite-size induced
complex~\cite{Aubry2006} multipliers). Nevertheless, near
the resonances with each relevant phonon mode, there will
be a portion of increasing energy-frequency dependence,
corresponding to breather instability, in
line with the corresponding
stability criterion of~\cite{holyoke}.

\begin{figure*}[htb]
\includegraphics[width=0.48\textwidth]{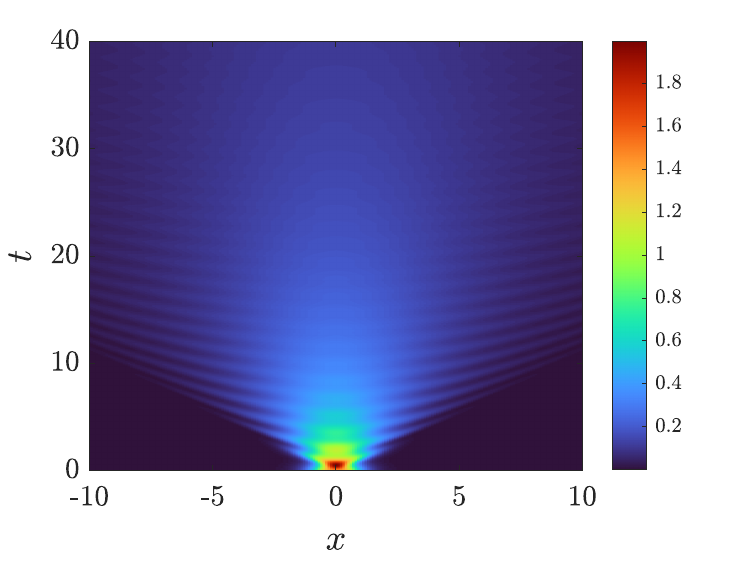}%
\hfill
\includegraphics[width=0.48\textwidth]{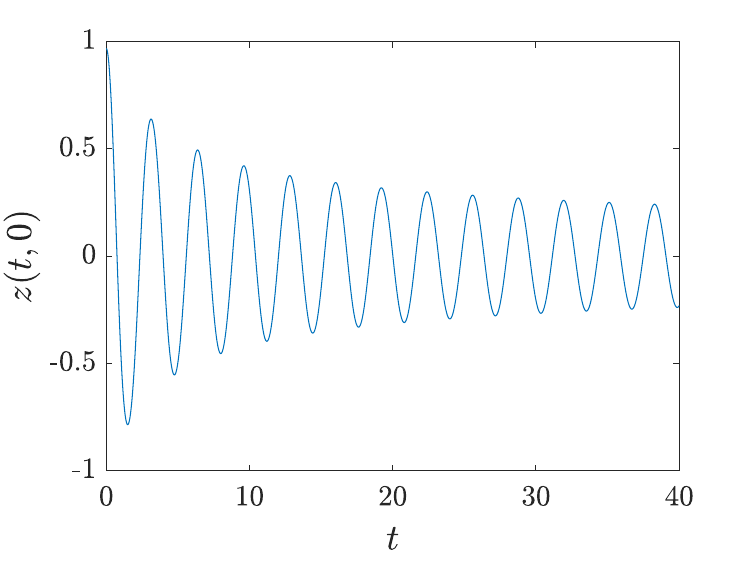}

\includegraphics[width=0.48\textwidth]{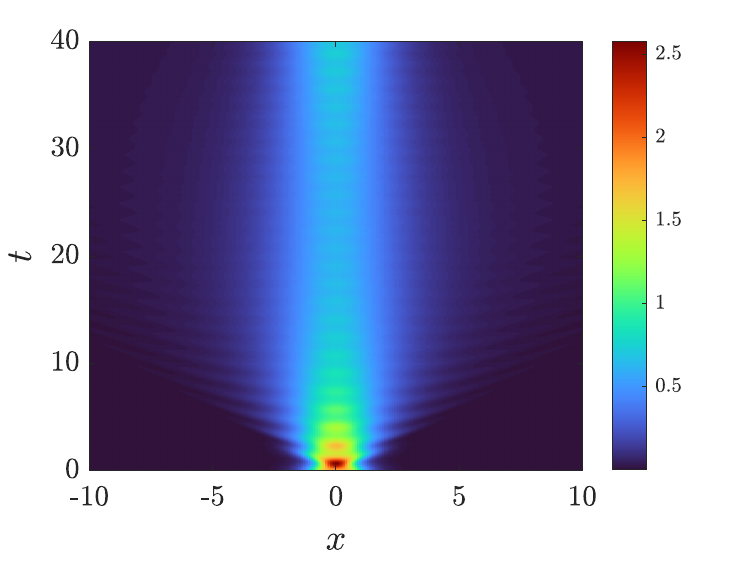}%
\hfill
\includegraphics[width=0.48\textwidth]{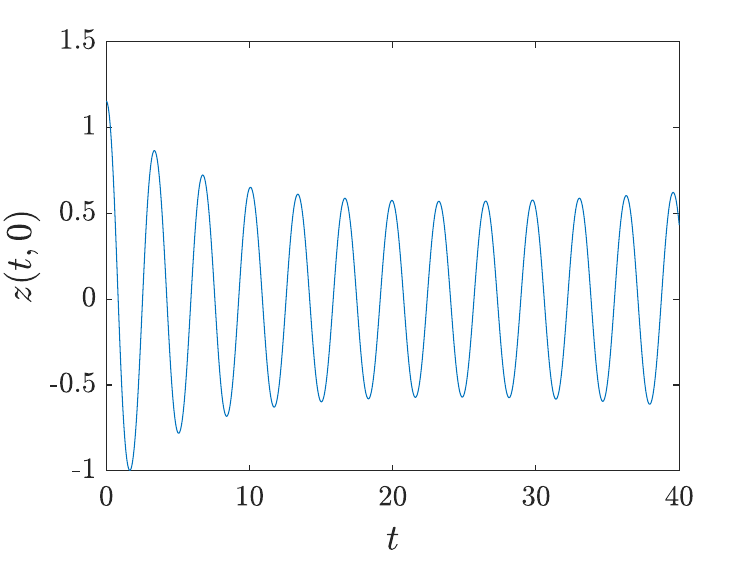}

\includegraphics[width=0.48\textwidth]{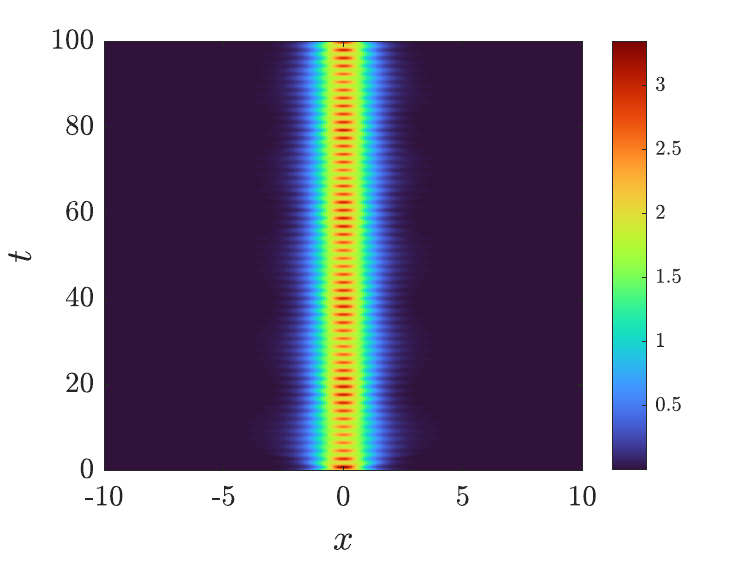}%
\hfill
\includegraphics[width=0.48\textwidth]{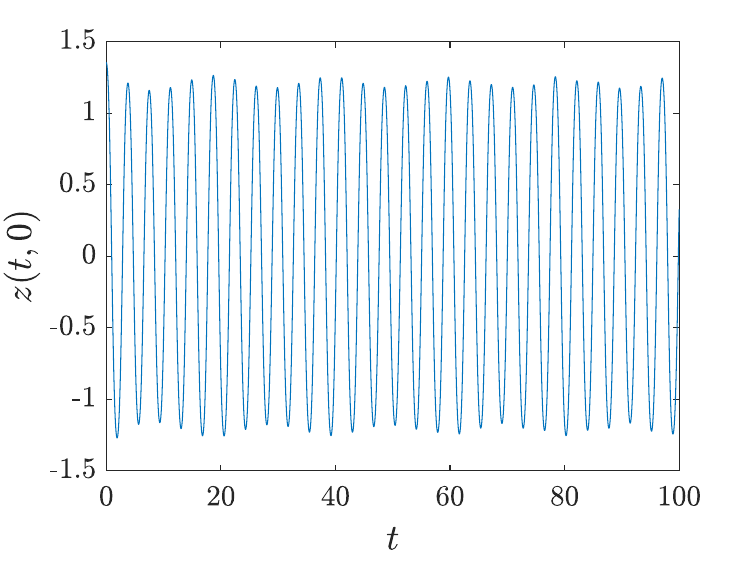}

\caption{(Left panels) Spatio-temporal evolution of the energy density for the KG equation using as initial condition a $(-,+)$ oscillon with frequency $\omega=0.4$ (top panels), $\omega=0.47$ (middle panels), and $\omega=0.515$ (bottom panels). The right panels show the evolution for $z(t,0)$, i.e., at $x=0$, for the cases in the left panels.}
\label{fig:simul4}
\end{figure*}

\begin{figure}[htb]
\includegraphics[width=0.48\textwidth]{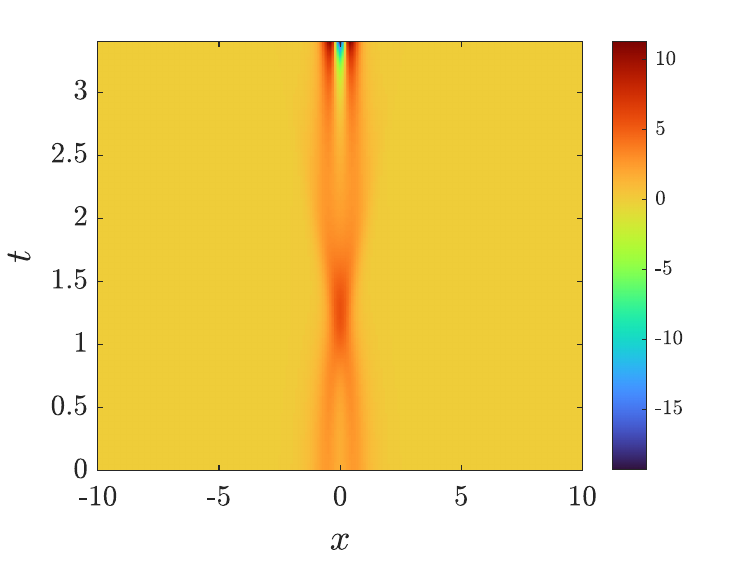}%
\hfill
\includegraphics[width=0.48\textwidth]{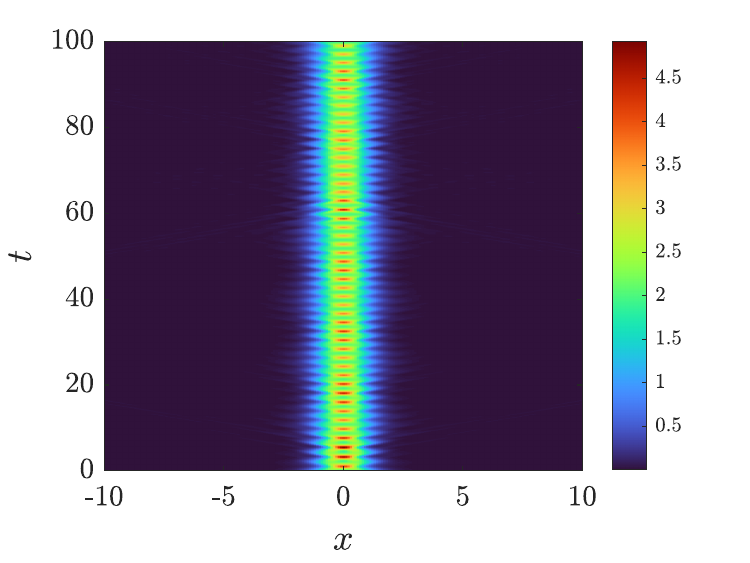}

\caption{Spatio-temporal evolution of the energy density for the KG equation using as initial condition a $(+,+)$ and a $(-,+)$ oscillon with frequency $\omega=0.53$. Left (right) panel corresponds to the case when the initial condition is taken from the black (red) curve of Fig.~\ref{fig:bif}.}
\label{fig:simul5}
\end{figure}

Finally, we have also considered the dynamics of $(-,+)$ and $(+,+)$ oscillons. As mentioned above, we have been unable to
systematically identify breathers pertaining to such steady
state solutions, hence, instead, in what follows, we have
focused on the initialization of the original KG PDE of
Eq.~(\ref{10}) with the solution of the steady state
problem of Eqs.~(\ref{30}). In the case of $(-,+)$ (see Fig.~\ref{fig:simul4}) we have observed that the oscillon generically disperses although it still oscillates as one can see from the panels showing $z(t,0)$; when the frequency increases,
the breather dispersion decreases and, if the growth rate is small enough, a robust quasi-periodic breather-like excitation 
is generated. As an example, one can observe that for the $(-,+)$ oscillon with $\omega=0.515$, resulting in a quasi-periodic breather with dominant frequency $1.684$ and secondary frequency $1.370$. 
For the $(+,+)$ oscillons, the dynamics typically lead to blow up, as one can see in the left panel of Fig.~\ref{fig:simul5}
(cf. with the right panel of the figure stemming from the
$(-,+)$ branch which leads again to quasi-periodic dynamics).

\section{Conclusions and Future Challenges}

In the present work we have revisited the study of oscillons
in one spatial and one temporal dimension in the class of
``flipped sign'' $\phi^4$ models in the spirit of the work
of~\cite{KosevichKovalev1975}. Motivated by the 
recent developments in the work of~\cite{baras_oscillon,Alexeeva2024}, we have provided
an alternative mathematical approach towards the study
of these oscillons, which, however, we have argued extends
well past this particular PDE. We have reduced the problem
into effectively a two-mode system, concerning the dynamics
of the first and third harmonic (given the nature of the
cubic nonlinearity). The steady states of the resulting PDE
system provide a starting point approximation towards
the identification of the full time-periodic oscillon
states of the original model. Here we have provided
a mathematical analysis, using the tools of index theory,
of dynamical systems (through the use of suitable Lyapunov
type functionals) and of nonlinear PDEs, developing 
quantities whose monotonicity change is tantamount to the
change of the steady state's stability.  We have then
identified the resulting states numerically and confirmed
our stability conclusions, as well as provided a detailed
numerical bifurcation analysis of such states, bearing numerous
unusual features. These included, e.g., a transcritical-like
bifurcation with symmetry but also other bifrcations 
(such as saddle-centers) between some of the relevant 
stationary states. Finally, we have compared these findings
with the full PDE model finding good agreement between the existence
and stability of one of the branches and the corresponding
PDE features. Other of the branches involved significant
contributions of higher harmonics and thus led to less meaningful
results at the level of full oscillons, although they involved
a considerable level of mathematical interest in their reduced
model analysis.

We believe that this approach sheds light on a mathematical
technique that can be more widely used for periodic solutions
in continuum (but possibly also in discrete) problems. 
Arguably, the resulting reduced PDE models and their steady states
are of interest in their own right, yet it seems relevant to
establish conditions under which the latter are more closely
connected to the original  problem and its
time-periodic waveforms. In that spirit,
such a technique could be used also for the higher-dimensional
study and analysis of oscillons that is of wide relevance
to cosmological problems~\cite{doi:10.1142/S0218271807009954}.
Such studies, both discrete and higher dimensional of
such time-periodic and potentially oscillon-type solutions
are currently under consideration and will be reported
in future publications. 

{\bf Conflict of interest statement:} The authors declare that they have no  competing financial interests or personal relationships that could have influenced the work reported in this paper. 

{\bf Data availability statement:}The data that support the findings of this study are mostly contained in the paper. Additional datasets can be made available from the corresponding author upon request. 

{ {\bf Acknowledgements:} We are grateful to 
Professor Igor Barashenkov for bringing this
important problem to our attention, as well
as for numerous constructive remarks on this
work. This material is based upon work supported by the U.S. National Science Foundation under the awards PHY-2110030, PHY-2408988, DMS-2204702 (PGK). J.C.-M. acknowledges support from the EU (ERDF program) through MICIU/AEI/10.13039/501100011033 under the project PID2022-143120OB-I00} 

\bibliographystyle{apsrev4-2}  
\bibliography{bibfile}         

\end{document}